\documentclass{IEEEtran}

\usepackage{subfigure}
\ifCLASSINFOpdf
\else
\fi
\ifCLASSOPTIONcompsoc
  \usepackage[caption=false,font=normalsize,labelfont=sf,textfont=sf]{subfig}
\else
  \usepackage[caption=false,font=footnotesize]{subfig}
\fi

\usepackage{amsmath}

\usepackage{url}

\usepackage{mathtools}
\usepackage[usenames, dvipsnames]{xcolor}
\usepackage[vlined,ruled,linesnumbered]{algorithm2e}
\usepackage{slashbox}

\newenvironment{definition}[1][Definition]{\begin{trivlist}
\item[\hskip \labelsep {\bfseries #1}]}{\end{trivlist}}

\newcommand{\qed}{\nobreak \ifvmode \relax \else
      \ifdim\lastskip<1.5em \hskip-\lastskip
      \hskip1.5em plus0em minus0.5em \fi \nobreak
      \vrule height0.75em width0.5em depth0.25em\fi}

\begin{document}

\title{Optimal Orchestration of Virtual Network Functions}
\author{Meihui Gao, 
Bernardetta Addis,
Mathieu Bouet, 
Stefano Secci, 
\IEEEmembership{Senior Member,~IEEE}
\thanks{A preliminary version of this paper appeared in~\cite{AddisEtAl2015}.}
\thanks{Meihui Gao and Bernardetta Addis are with  LORIA, UMR 7503, Universit\'e de Lorraine, France.
Email: \{bernardetta.addis,meihui.gao\}@loria.fr.}
\thanks{Mathieu Bouet is with Thales Communications \& Security.
Email: mathieu.bouet@thalesgroup.com.}
\thanks{Stefano Secci is with Sorbonne Universit\'es, UPMC Univ Paris 06, UMR 7606, LIP6. Email: stefano.secci@upmc.fr.}}

\maketitle
\thispagestyle{empty}
\begin{abstract}
The emergence of Network Functions Virtualization (NFV) is bringing a set of novel algorithmic challenges in the operation of communication networks. NFV introduces volatility in the management of network functions, which can be dynamically orchestrated, i.e., placed, resized, etc. Virtual Network Functions (VNFs) can belong to VNF chains, where nodes in a chain can serve multiple demands coming from the network edges. 
In this paper, we formally define the VNF placement and routing (VNF-PR) problem, proposing a versatile linear programming formulation that is able to accommodate specific features and constraints of NFV infrastructures, and that is substantially different from existing virtual network embedding formulations in the state of the art. We also design a math-heuristic able to scale with multiple objectives and large instances.
By extensive simulations, we draw conclusions on the trade-off achievable between classical traffic engineering (TE) and NFV infrastructure efficiency goals, evaluating both Internet access and Virtual Private Network (VPN) demands. We do also quantitatively compare the performance of our VNF-PR heuristic with the classical Virtual Network Embedding (VNE) approach proposed for NFV orchestration, showing the computational differences, and how  our approach can provide a more stable and closer-to-optimum solution.
\end{abstract}


\IEEEpeerreviewmaketitle

\section{Introduction}
\label{intro}

After about ten years of fundamental research on network virtualization and virtual network embedding, the virtualization of network functions is becoming a reality thanks to huge investments being made by telecommunication providers, cloud providers and vendors.

The breaking point sits in 2012, when calls for experimentation and deployment of what was coined as \lq\lq Network Functions Virtualization (NFV)\rq\rq~\cite{chiosi2012network} lead to the creation of an NFV industry research group at the European Telecommunications Standards Institute (ETSI)~\cite{NFVWP}.
Since then, applied researches and developments have accelerated  investments, hence preliminary  prototypes were demonstrated and deployed (leading to commercialization in some cases) since late 2014~\cite{NFV-commag15}.

With NFV, the attention of network virtualization research is now focusing on key aspects of NFV systems that were either not considered relevant or not conceived before industry effort at Standards Developing Organizations (SDOs). Key aspects that are worth being mentioned are the:

\begin{itemize}
\item   NFV service chaining~\cite{quinn2015network} provisioning, i.e., the problem of allowing a traffic flow passing through a pre-computed or dynamically computed list of VNF nodes, possibly accounting for the fact that VNF nodes can be placed at, and migrated across, virtualization clusters as a function of demand assignment to existing VNF chains or sub-chains;

\item  ingress/egress bit-rate variations at VNFs, due to specific VNF operations (such as compression as in coding, decompression as in tunneling);

\item  VNF processing and forwarding latency as an orchestration parameter. It can indeed be exponential with the traffic load on the VNF, or constant up to a maximum board if computation offloading solutions, such as direct memory access bypassing the hypervisor (as done with Intel/6WIND Data-Plane Development Kit~\cite{dpdk}), or similar other \lq fastpath\rq \ solutions are present. 
\end{itemize}

We could not identify a work in the state of the art jointly taking these aspects all together into account. As hereafter resumed, most of the approaches rely on heuristic algorithms. A recent study~\cite{Fischer2016} evaluate some of them highlighting that they may come at a \lq\lq Revenue/Cost\rq\rq \ ratio of 50\%, i.e., twice as many resources were consumed than demands realized by heuristic approaches, which suggests that there is significant optimization potential to achieve in the area, despite the high number of research papers on VNF orchestration. In this paper, our goal is to fill the cost efficiency gap mentioned in~\cite{Fischer2016} under reasonable execution time.

ETSI is de-facto the reference SDO for the NFV high-level functional architecture specification.
High-level means that its identified role is the specification of the main functional blocks, their architecture and inter-relationship, whose implementation elements could then be precisely addressed by other SDOs.
ETSI specifies three components~\cite{NFVGS002} for the NFV architecture: Virtual Network Functions (VNFs); NFV Infrastructure (NFVI), including  the elements needed to run VNFs such as the hypervisor node and the virtualization clusters; MANagement and Orchestration (MANO), handling the operations needed to run, migrate, optimize VNF nodes and chains, possibly in coordination with transport network orchestrators.

\begin{figure}[t]
\centering 
\begin{subfigure}[Traditional CPE]{\includegraphics[width=88mm]{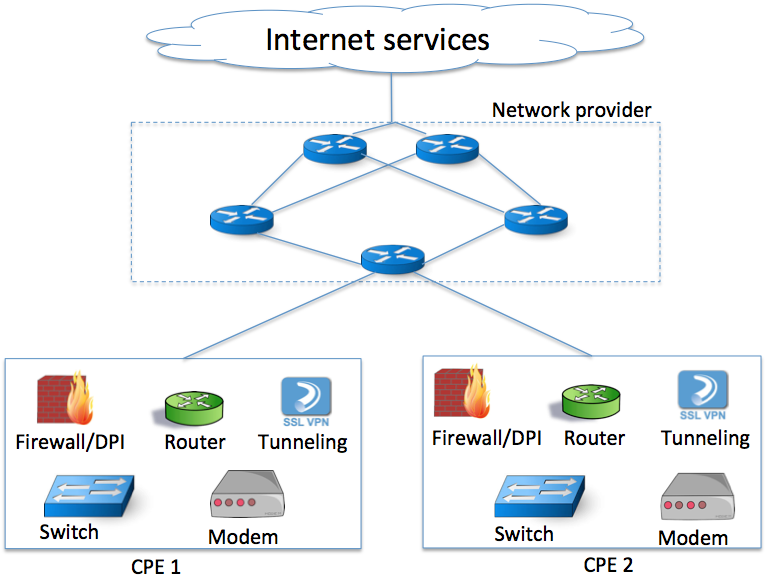}}
\label{vcpe1} 
\end{subfigure}
\begin{subfigure}[vCPE]{\includegraphics[width=70mm]{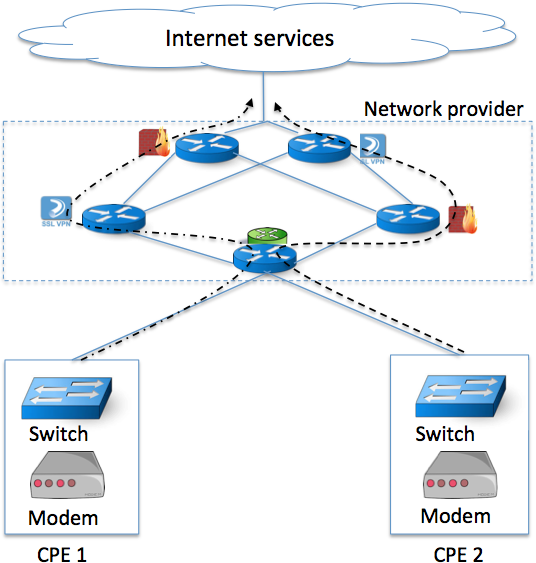}}
\caption{Traditional Customer Premises Equipment (CPE) compared to virtualized CPE (vCPE) with VNF chaining.}
\label{vcpe2} 
\end{subfigure}
\label{vcpe}
\end{figure}

MANO procedures come therefore to support the economies of scale of NFV, so that physical NFVI virtualization resources (servers and clusters) dedicated to NFV operations are used efficiently with respect to both NFVI operators and edge users.
A promising NFV use-case~\cite{NFVGS001} for carrier networks is the virtual Customer Premises Equipment (vCPE) that simplifies the CPE equipment by means of virtualized individual network functions placed at access and aggregation network locations, as depicted in Fig.~\ref{vcpe}. There are also other promising use-cases like the virtualization of the Evolved Packet Core (EPC) cluster in cellular core networks~\cite{Basta2014, Hawilo2014}, and the virtualization of cellular base stations~\cite{Wu2015}.

MANO operations are many and range from the placement and instantiation of VNFs to better meet user's demands to the chaining and routing of VNF chains over a transport network disposing of multiple NFVI locations. Part of the orchestration decision can also be the configuration of the VNFs to share them among active demands, while meeting common Traffic Engineering (TE) objectives in IP transport networks as well as novel NFV efficiency goals such as the minimization of the number of VNF instances to install. In this context, the paper contribution is as follows:
\begin{itemize}
\item we define and formulate via mathematical programming the VNF Placement and Routing (VNF-PR) optimization problem, including compression/decompression constraints and two forwarding latency regimes (with and without fastpath), under both TE and NFV objectives.
\item we compare the VNF-PR approach to the legacy  Virtual Network Embedding (VNE) approach, qualitatively and quantitatively;
\item we design a math-heuristic approach allowing us to run experiments also for large instances of the problem within an acceptable execution time.
\item we evaluate our solution by extensive simulations. 
We draw considerations on NFV deployment strategies.
\end{itemize}

The paper is organized as follows. Section \ref{Background} presents the state of the art on NFV orchestration. Section \ref{netm} describes the network model and the Mixed Integer Linear Programming (MILP) formulation. Analysis and discussion of optimization results are given in Section \ref{results}. Section \ref{concl} concludes the paper.

\section{Background}
\label{Background}
Network virtualization research was first driven by the convergence of computation, storage and network in cloud computing. A large number of works in the literature address the optimization of Virtual Machines (VM) placement with respect to, for example, server load balancing or energy saving~\cite{bobroff2007dynamic,chaisiri2009optimal}. 
Virtualizing the network between VMs is also a problem addressed in the area, defined as the Virtual Network Embedding (VNE) problem of mapping a set of logical graphs of interconnected VMs on a substrate graph~\cite{Fischer2013}.

In NFV, network functions that were once run by hardware-based middleboxes~\cite{Sherry2012} are now meant to be virtualized as VNFs.
VNFs can be chained together to provide a specific service, also known as service/VNF chaining. Service providers can deploy specific service chains to provide network service demands that requested by clients.

Preliminary works on NFV orchestration tend to solve the NFV orchestration problem as a VNE problem, which treats virtual network requests as logical graphs to be embedded into a substrate network. This is for example the case of~\cite{guerzoni2014novel}. VNFs are treated as normal VMs, mapped on a network of VM containers, which are interconnected via physical links that host logical links of virtual network demands. 
Similarly, authors in~\cite{mehraghdam2014specifying} propose a VNF chain placement that combines location-routing problems and VNE problems, solving first the placement and then the chaining.
In~\cite{moens2014vnf} the authors decouple the legacy VNE problem into two embedding problems: VM embedding and service chain embedding, where a service chain is embedded on VMs, and each VM on physical servers. Each service chain has specific requirements as notably an end-to-end latency requirement. 

The placement and routing of VNFs is a problem fundamentally different from the VNE problem. As in VNE, virtual network nodes need to be placed in an underlying physical infrastructure. However, differently from VNE, in VNF placement and routing: (i) the demand is not a multipoint-to-multipoint network connection request, but a point-to-point source-destination flow routing demand, 
and (ii) specific aspects of NFV such as forwarding latency behavior, ingress/egress bit-rate changes, and chaining are not addressed in VNE. Their inclusion would further increase the VNE time complexity (for instance, in~\cite{Botero2012} forwarding latency is considered by adding \lq hidden nodes\rq hence largely increasing the spatial and time complexities). In this sense VNF placement and routing problem is closer to a facility location problem, whereas VNE is closer to a mapping problem.

We argue in this paper that the appropriate way to deal with NFV MANO decision problems~\cite{Bo2015, Mijumbi2016} is to define the VNF Placement and Routing (VNF-PR) problem directly tailored to the NFV environment, for the sake of time complexity, modeling precision and practical usability. 
This is also the approach adopted by a few papers in the literature~\cite{netsoftVNFchain,Cohen2015,Luizelli2015}. 
In~\cite{netsoftVNFchain} the authors consider the online orchestration of VNFs, modeling it as a scheduling problem of VNFs and proposing heuristics to scale with the online nature of the framework.
In~\cite{Cohen2015} the authors study a bi-criteria approximation algorithm for the cost minimization objective function as well as for the nodes size constraints.
In~\cite{Luizelli2015} the authors propose a formulation of the VNF placement and chaining problem and an Integer Linear Programming (ILP) model to solve it. Additionally, to cope with large infrastructures, they introduce a binary search procedure for efficiently guiding the ILP solver towards feasible, near-optimal solutions. Different to~\cite{netsoftVNFchain,Cohen2015,Luizelli2015}, in this paper, we focus on providing a more generic formulation to the VNF placement and routing problem. Apart from chaining with VNF ordering guarantees, we also capture and investigate the practical feature of traffic flow compression and decompression that could be imposed by the VNFs along the traffic route.

A common approach is to rely on graph properties to find a better utilization of the limited resources while serving a larger set of demands.
In~\cite{bouetcost} the specific Deep Packet Inspection (DPI) VNF node placement problem (without chaining) is targeted, with a formal definition of the problem and a greedy heuristic algorithm to solve it.
In~\cite{Kuo2016} the authors propose a heuristic to place and chain a maximum number of VNFs under capacity limitation; a linear programming approach is formulated to iterate the $k$-shortest paths computation for each VNF chain and choose the one that satisfies the maximal length and number of reused VNFs.
In comparison, our proposal considers not only the efficiency of resources utilization, but also the quality of service provisioning (for instance, the traffic forwarding latency). Moreover, we discuss the trade-off between resources efficiency goals and network traffic engineering goals.

Recently, game theoretic approaches are also considered: in~\cite{Obadia2016} authors propose a heuristic based on routing games; in~\cite{Elias2015} authors propose a distributed dynamic pricing approach to allocate demands to already placed VNF instances, with convex congestion functions for both links and VNFs to control congestion.

Finally, the VNF placement and routing problem has also been addressed in specific contexts: wireless local area networks~\cite{Riggio2015} and optical networks~\cite{xia2014network}.
A comprehensive survey on NFV resource allocation and chaining was recently published in~\cite{Herrera2016}.

Our paper takes inspiration from these early works, yet goes beyond being more generic and integrating the specific features of NFV environments mentioned in the introduction and formalized in the following.

\section{Network Model}
\label{netm}

We provide in the following a problem statement, its mathematical programming formulation, and a description of its possible customization alternatives.

\subsection{Problem statement}

\begin{definition} Virtual Network Function Placement and Routing (VNF-PR) Problem\\
The network is represented by a graph $G(N, A)$, where $N$ is
the set of switching nodes, $A$ represents the possible directional
connections between nodes. The router $i \in N$ and its associated
NFVI  cluster are represented by the same node; this
choice allows to keep the size of the graph limited and reduces
the computational effort. We represent with $N_v \subset N$ the set
of nodes $N$ disposing of NFVI server clusters. We consider
a set of demands $D$, each demand $k \in D$ is characterized
by a source $o_k$, a destination $t_k$, a nominal bandwidth $b_k$
(statistically representative for demand $k$), and a sequence of
VNFs of different types, that must serve the demand (and
therefore must be traversed by the demand). For each VNF a
single VM is reserved, therefore we can equivalently speak of
allocating a VM or a VNF on a NFVI cluster, meaning that
we are reserving the necessary resources (e.g., CPU, RAM) to host
a VM running a VNF.
The VNF-PR optimization problem is to find:
\begin{itemize}
\item the optimal placement of VNF nodes over NFVI clusters;
\item the optimal routing for demands and their assignment to
VNF node chains.
\end{itemize}
subject to:
\begin{itemize}
\item link capacity constraints;
\item NFVI cluster capacity constraints;
\item VNF flow compression/decompression constraints;
\item VNF forwarding latency constraints;
\item VNF node sharing constraints;
\item VNF chain (total or partial) order for each demand.
\end{itemize}
\end{definition}

\begin{figure}[t]
\centering
\includegraphics[width=80mm]{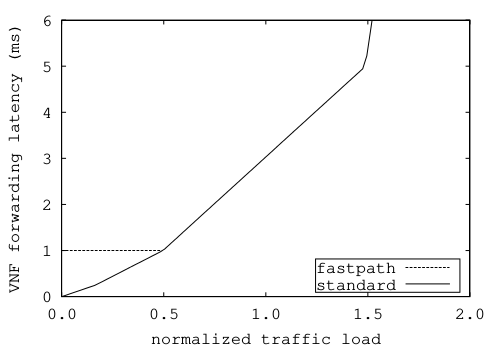}
\caption{Example of VNF forwarding latency profiles.}
\label{latency}
\end{figure}

The optimization objective should contain both network-level and NFVI-level performance metrics. In our network
model, we propose as network-level metric a classical TE metric, i.e., the minimization of the maximum link utilization. As
NFVI-level metric we propose the minimization of allocated
computing resources. Furthermore, we assume that:
\begin{itemize}
\item Multiple VNFs of the same type (i.e., same functionality)
can be allocated on the same node, but each demand
cannot split its flow on multiple VNFs of the same type.
\item  The VNF computing resource consumption can be expressed in terms of live memory (e.g., RAM) and Computing Processing Units (CPUs), yet the model shall be
versatile enough to integrate other computing resources.
\item  Latency introduced by a VNF can follow one among the
two following regimes (as represented in Fig.~\ref{latency}):
\begin{itemize}
\item \textit{Standard}: VNFs bufferize traffic at input and output
virtual and physical network interfaces such that the forwarding latency can be considered as a convex piece-wise
linear function of the aggregate bit-rate at the VNF,
due to increased buffer utilization and packet loss
as the bitrate grows as shown in~\cite{dpdk,dpdk2}. This is
the case of default VNFs functioning with standard
kernel and hypervisor buffers and sockets.
\item \textit{Fastpath}: VNFs use optimally dimensioned and relatively small buffers, and decrease the number of
times packets are copied in memory, so that the
forwarding latency is constant up to a maximum
aggregate bit-rate after which packets are dropped (e.g., this happens for Intel/6WIND DPDK fastpath
solutions~\cite{dpdk}).
\end{itemize}
Fig.~\ref{latency} gives examples of forwarding latency profiles for
the two cases.
\item  For each demand and NFVI cluster, only one compression/decompression VNF can be installed. This allows us
to keep the execution time at acceptable levels, without
reducing excessively the VNF placement alternatives.
This assumption can be relaxed at the cost of working on
an extended graph, and therefore increasing the computational time of the algorithm.
\end{itemize}

{
\footnotesize
\begin{table} 
\caption{General mathematical notations} 
\begin{tabular}{c|l}
\multicolumn{2}{c}{Sets}\\
\hline
$N$ & all nodes\\
$N_v \subseteq N$  & nodes equipped with a NFVI cluster\\
$A \subseteq N \times N$ & all arcs (links)\\
$D$   & demands\\
\hline
$R$   & resource types (CPU, RAM, ...)\\
$F$   & VNF types\\
\hline
\multicolumn{2}{c}{Parameters}\\
\hline
\multicolumn{2}{c}{network parameters}\\
\hline
$\gamma_{ij}$ & link capacity\\
$\Gamma_{ir}$ &  capacity of node $i \in N_v$ in terms of resource $r \in R$\\
\hline
\multicolumn{2}{c}{demand parameters}\\
\hline
$o_k$ & origin of demand $k \in D$\\
$t_k$ & destination of demand $k \in D$\\
$b_k$  & nominal bandwidth of demand $k \in D$\\
$\mathit{m}_{k}^{f}$ & $1$ if demand $k \in D$ requests VNF of type $f \in F$\\
$\mathit{s}_{k}^{f}$ & order coefficient for VNF $f$ requested by demand $k$\\
\hline
\multicolumn{2}{c}{VNF/VM parameters}\\
\hline
$\mathit{rr}_{r}$ & demand of resource $r \in R$ for a VM\\
$c_i^f$ & maximum number of copies of VNF $f$ on node $i$\\
\hline
\multicolumn{2}{c}{Variables}\\
\hline
\multicolumn{2}{c}{binary variables}\\
\hline
$x_{ij}^k$    & $1$ if arc $(i,j)$ is used by demand $k \in D$\\
$z_{ik}^{fn}$ & $1$ if demand $k \in D $ uses copy $n$-th of VNF\\
			  &of type $f \in F$ placed on node $i \in N_v$\\
$y_{i}^{fn}$  & $1$ if $n$-th copy of a VNF $f$ is assigned\\
			  &to node $i \in N_v$\\
$w_{ik}^{f}$ & 1 if demand $k$ uses VNF $f$ on node $i \in N_v$\\
\hline
\multicolumn{2}{c}{continuous variables}\\
\hline
$U \ge 0$ & maximum link utilization\\
$\pi_{ik} \ge 0$ & position of node $i$ in the path used by demand $k$\\
\hline
\end{tabular}\label{basemodel}
\end{table}
}

\subsection{Mathematical formulation}

We first introduce a basic model that does not take into
account latency limitations and compression/decompression
features. The reason of this choice is twofold. First, it
allows a clearer explanation of the model and a step by
step introduction of the technicalities that allow us to keep
the model linear; we recall that already without these two
features, the model is a combination of a network design
and a facility location. Second, in the algorithmic phase we
used for solving (exactly) the model, we solve a sequence of
models with increasing complexity (basic model, with latency,
with compression/decompression), therefore, this presentation
allows to put in evidence the peculiarities of each model.
\subsubsection{Basic VNF-PR model}
Table~\ref{basemodel} reports the mathematical notations used in the following Mixed Integer Linear
Programming (MILP) that represents the basic formulation
of the VNF-PR problem. We use four families of binary
variables: $x_{ij}^k$ represents the per-demand link utilization, hence
the path used by the demand; $y_i^{f n}$ represents the allocation of copy $n$ of a VNF of type $f$ on a given node; $w_{ik}^f$ represents the
assignment of a given demand to a VNF and $z_{ik}^{fn}$
represents the assignment of a given demand to a specific copy of a VNF.
The continuous variable $\pi_{ik}$
is used to represent the position of
node $i$ in the path used to route demand $k$. This family of
variables is necessary to impose (total or partial) order in the
VNF chain. As mentioned before, we consider two objective
functions:
\begin{itemize}
\item TE goal: minimize the maximum network link utilization:
\begin{equation}
\min U
\label{objTE}
\end{equation}

\item NFV goal: minimize number of cores (CPU) used by the instantiated VNFs: 
\begin{equation}
\min \sum_{i\in N_v} \sum_{f\in F} \sum_{n \in 1..c_i^f} rr_{CPU} \ y_{i}^{fn}
\label{objNFV}
\end{equation}

\end{itemize}

The former objective allows taking into consideration the
inherent fluctuations related to Internet traffic and therefore
minimizing the risk of sudden bottleneck on network links.
The latter assumes the fact that today the first modular
cost in virtualization servers, especially in terms of energy
consumption, is the CPU. We now present the constraints.

Single path flow balance constraints:

\begin{equation}
\sum_{j: (i,j) \in A} x_{ij}^k - \sum_{j: (j,i) \in A} x_{ji}^k = \left \{
   \begin{array}{r c l}
      1  \ \text{if} \ i=o_k\\
      -1 \ \text{if} \ i=t_k\\
      0  \ \text{otherwise}
   \end{array}
   \right . \ \forall k \in D, \forall i \in N
\label{eq:singlepath}
\end{equation}

Utilization rate constraints:
\begin{equation}
\sum_{k \in D} b_k x_{ij}^k \le U\gamma_{ij} \quad \forall (i,j) \in A\label{eq:linkutil}\\
\end{equation}

Node resource capacity (VNF utilization) constraints:
\begin{equation}
\sum_{f \in F} \sum_{n \in 1..c_i^f} \mathit{rr}_{r} y_{i}^{fn} \le \Gamma_{ir} \quad \forall i \in N_v \label{eq:servercap}\\
\end{equation}

Each demand uses exactly one VNF of each required type:
\begin{equation}
\sum_{i \in N_v} \sum_{n \in  1..c_i^f} z_{ik}^{fn} = 1 \quad \forall k \in D, f \in F : \mathit{m}_{k}^{f}=1 \label{eq:atleastNVF}\\
\end{equation}

Constraints~\eqref{eq:zycohe}-\eqref{eq:zwcohe} are consistency constraints among binary
variables. 
A VNF can be used only if present, for a given node:
\begin{equation}
z_{ik}^{fn} \le y_{i}^{fn} \quad \forall k \in D, i \in N_v, f \in F, n \in  1..c_i^f \label{eq:zycohe}\\
\end{equation}

If a demand does not pass by a VNF, it cannot use it:
\begin{equation}
z_{ik}^{fn} \le \sum_{j: (j,i) \in A} x_{ji}^k \quad \forall k \in D, i \in N_v, f \in F: \mathit{m}_{k}^{f}=1 \label{eq:zxcohe}\\
\end{equation}

Auxiliary variables for ensuring consistency:
\begin{equation}
\sum_{n \in  1..c_i^f } z_{ik}^{fn} = w_{ik}^f, \quad \forall k \in D, i \in N_v, f \in F \label{eq:zwcohe}\\
\end{equation}

Finally, we introduce constraints to avoid unfeasible routing and to impose the VNF chain order:

Preventing the formation of isolated cycles:
\begin{equation}
\pi_{jk} \ge \pi_{ik} + x_{ij}^k - |N_v|(1-x_{ij}^k) \quad \forall k \in D, (i,j) \in A \label{eq:nocycles}
\end{equation}
Imposing an order for virtual functions:
\begin{eqnarray}
\pi_{jk} \ge & \pi_{ik} - (|N_v|+1)(2-w_{ik}^{f_1}-w_{ik}^{f_2}) \quad \forall k \in D,\nonumber\\ 
		     & \forall i,j \in N_v, f_1,f_2 \in F: s_k^{f_2} \geq s_k^{f_1}\label{eq:order}
\end{eqnarray}

If we consider flow balance constraints~\eqref{eq:singlepath} and link capacity
constraints~\eqref{eq:linkutil}, for each demand, a selection of arcs forming a
path plus an isolated cycle can be a feasible solution. In pure
routing problems, these solutions are equivalent to the solution
where routing variables along the cycle are removed and only
the one along the path are kept. In fact, both constraints~\eqref{eq:singlepath} and
constraints~\eqref{eq:linkutil} will be valid for this new solution. Our problem
integrates routing features within a facility location problem,
therefore such solutions cannot always be transformed in a
simple path simply removing the cycle. In fact, if a facility
(VNF) used by the demand is located on the cycle, removing
the cycle will produce an unfeasible solution. Therefore, it is
necessary to remove such solutions directly in the model, to
this aim we introduced constraints~\eqref{eq:nocycles}, inspired by traveling
salesman problems tour elimination constraints~\cite{Papadimitriou98}. Variable
$\pi_{ik}$ represents the order of node i along the path serving
demand $k$, therefore if arc $(i,j)$ exists, then $\pi_{jk}$ will be at
least $\pi_{ik}$ plus $1$. On the other side, if arc $(i,j)$ does not exist
($x_{ij}^k = 0$) then the constraint is not active: $\pi_{ik}$ is
always smaller than $|N_v|$, as a path can contain at most all
the nodes in the graph. In this way, only solutions containing
simple paths are allowed. These variables are also used in
Equation~\eqref{eq:order} to allow imposing an order on VNFs along the
route of a demand $k$. They impose that if demand $k$ uses VNF
$f_1$ located on node $i$ and its VNF successor $f_2$ ($s_k^{f_2} \geq s_k^{f_1}$ ) that
is located on node $j$, then in the routing path of demand $k$
node $i$ must be precede node $j$.

\subsubsection{VNF forwarding latency} 
{
\footnotesize
\begin{table} 
\caption{Notations to model forwarding latency} 
\begin{tabular}{c|l}
\hline
\multicolumn{2}{c}{}\\
\multicolumn{2}{c}{Parameters}\\
\hline
$L$ 		   & maximum allowed latency for a demand\\
$\lambda_{ij}$ &latency introduced by link $(i,j) \in A$\\
\hline
\multicolumn{2}{c}{\textit{standard} latency model}\\
\hline
$g_{j}^f(b)$ & $j$-th component of the linearized latency function\\
			 &for VNF $f$ and aggregated bandwidth $b$\\
$n_g$        &number of piece-wise components of lin. latency function\\
\hline
\multicolumn{2}{c}{\textit{fastpath} latency model}\\
\hline
$\bar{l}^f$  & latency introduced by VNF $f$\\
$B_{max}^f$  &maximum allowed bandwidth to traverse VNF $f$\\
\hline
\multicolumn{2}{c}{}\\
\multicolumn{2}{c}{Variables}\\
\hline
$l_{ik}^f \ge 0$ & latency that demand $k \in D$ incurs using VNF $f$\\
				   & on node $i$ of type $f \in F$ hosted by node $i \in N_v$\\
\hline
\end{tabular}\label{baselat}
\end{table}
}

We impose that for each demand $k \in D$
a maximum latency $L$ is allowed, to guarantee some level
of QoS. Latency depends on two components: link latency,
represented by a parameter $\lambda_{ij}$ for each arc $(i,j)$, and VNF
latency. VNF latency depends on the used model of latency
(\textit{standard} or \textit{fastpath}). To keep the notation as uniform as
possible, we introduce an additional variable $l_{ik}^f$
to represent
the latency experienced by demand $k$ traversing VNF $f$ located
on node $i$. Therefore we get a set of constraints common to
both models limiting the overall latency:
\begin{equation}
\sum_{(i,j) \in A} \lambda_{ij} x_{ij}^{k}+\sum_{i \in N_v}\sum_{f \in F} l_{ik}^{f} \le L \quad \forall k \in D\label{eq:demandlatency}
\end{equation}
A set of constraints depending on the chosen latency model
allows to calculate the value of variable $l_{ik}^f$.
\begin{itemize}
\item \textit{Standard}: the latency introduced on demand $k$ for using
VNF $f$ depends on the overall traffic traversing the VNF
(its own and the one of others demands). Let call $g_j^f(\cdot)$
the $j$-th component of of the piece-wise linearization
of the latency function for VNF of type $f$, then we get:
\begin{eqnarray}
l_{ik}^{f} \ge g_{j}^{f}(\sum_{d \in D} b_k z_{id}^{fn}) - L(1-z_{ik}^{fn})\nonumber\\ 
\quad \forall k \in D, i \in N_v, f \in F, n \in 1..c_i^f, j \in 1..n_g \label{eq:nodelatency-st}
\end{eqnarray}

We can observe that constraint~\eqref{eq:nodelatency-st} is active only when
the demand uses copy $n$ of VNF $f$ on node $i$ ($z_{ik}^{fn} = 1$).
Otherwise, as overall latency is limited by $L$ (constraint~\eqref{eq:demandlatency}), any term $g_j^f(\cdot)$ must be smaller than $L$, and therefore, the constraint is a redundant constraint
($l_{ik} \ge 0$). We can observe that, even if in the \textit{standard}
latency model there is no limit on the allowed bandwidth,
constraint~\eqref{eq:demandlatency}, limiting the overall latency, and
as a consequence latency on VNFs, imposes an implicit
limit on the allowed bandwidth.
\item \textit{Fastpath}: the latency is fixed, but a limit in the total traffic
that a VNF can support is imposed. Therefore we get the
following two sets of constraints:
\begin{equation}
l_{ik}^f = \bar{l}^f \quad \forall k \in D, i \in N_v, f \in F \label{eq:nodelatency-fp1}
\end{equation}
\begin{equation}
\sum_{d \in D} b_d z_{id}^{fn} \le B_{max}^f \quad \forall i \in N_v, f \in F, n \in 1..c_i^f \label{eq:nodelatency-fp2}
\end{equation}
We can observe that constraints~\eqref{eq:nodelatency-fp1} can be substituted
directly in constraints~\eqref{eq:demandlatency}.
\end{itemize}

\subsubsection{Bit-rate compression/decompression}
{
\footnotesize
\begin{table} 
\caption{Notations to model bit-rate variations} 
\begin{tabular}{c|l}
\hline
\multicolumn{2}{c}{}\\
\multicolumn{2}{c}{Sets}\\
\hline
$N_a$ 		   & access nodes\\
$N'_a$ 		   & duplication of access nodes, where demands are located\\
\hline
\multicolumn{2}{c}{}\\
\multicolumn{2}{c}{Parameters}\\
\hline
$\mu_f$ 		   & compression/decompression factor for VNF $f \in F$\\
\hline
$b_k^{min}$        &minimal bandwidth of demand $k \in D$\\
$b_k^{max}$        &maximal bandwidth of demand $k \in D$\\
$M_i$              &maximum traffic volume that can be switched by node $i$\\               
\hline
\multicolumn{2}{c}{}\\
\multicolumn{2}{c}{Variables}\\
\hline
$\phi_{ij}^k \ge 0$ & flow for demand $k \in D$ on arc $(i,j)$\\
$\psi_{ik}^{fn} \ge 0$ & flow for demand $k \in D$ entering node $i$\\
					   & and using copy $n$ of VNF $f \in F$\\
\hline
\end{tabular}\label{basecd}
\end{table}
}

To introduce the
possibility of compressing/decompressing flows for some
VNFs, we need some modifications to our model description.
We introduce a compression/decompression parameter $\mu_f$ for
each type of VNFs, $\mu_f > 1$ means that a decompression
is performed by VNF $f$. When a demand pass through a
VNF with $\mu_f \neq 1$, its bandwidth changes, therefore knowing
just the routing ($x$ variables) is not enough to determine
the overall flow along an arc. For this reason we introduce
variable $\phi_{ij}^k$ that represents  explicitly the flow on arc $(i,j)$
for demand $k$. Another consequence is that the classical flow
balance equations are not anymore valid. To extend the model
without introducing an excess of complexity, we work under
the assumption that given a node $i$, and a demand $k$, such
demand uses at most a VNF $f$ with a factor of compression/decompression ($\mu_f \neq 1$). We work on an extended graph
to distinguish between access nodes (origin/destination nodes)
and NFVI nodes (remind that with the basic model we collapsed NFVI nodes on router nodes). Each access node $i$ is duplicated in a node
$i'$. Arc $(i,i')$ will be added and all arcs $(i,j)$ originating from
access node $i$ will be transformed in arcs $(i',j)$. Therefore, the routing functionality is on node $i$ and the NFVI functionality can be allocated on node $i'$. Furthermore, we add variable $\phi_{ik}^{fn}$, that represents the flow of demand $k$
entering node $i$ and using the copy $n$ of the VNF of type $f$.
If a demand passes through a VNF with a factor of compression/decompression $\mu_f$, then the out-flow of the node is proportional to the in-flow:
\begin{equation}
\sum_{j \in N: (i,j) \in A} \phi_{ij}^k= \mu_{f}\sum_{j \in N: (j,i) \in A} \phi_{ji}^k\nonumber
\end{equation}
or equivalently: 
\begin{eqnarray}
\sum_{j \in N: (i,j) \in A} \phi_{ij}^k - \sum_{j \in N: (j,i) \in A} \phi_{ji}^k =
\sum_{j \in N: (j,i) \in A} (\mu_{f}-1) \phi_{ji}^k\nonumber
\end{eqnarray}
This equation is valid only if demand $k$ uses a copy $n$ of
VNF $f$ on given node $i$ (remind that latency depends on the
bandwidth passing throw a single copy). Therefore, to obtain a valid equation, we have to write:
\begin{eqnarray}
\sum_{j \in N: (i,j) \in A} \phi_{ij}^k - \sum_{j \in N: (j,i) \in A} \phi_{ji}^k = \nonumber\\
\sum_{j \in N: (j,i) \in A} \phi_{ji}^k \sum_{n \in 1..c_{i}^{f}}(\mu_{f}-1)z_{ik}^{fn}\nonumber
\end{eqnarray}
when $\sum_{n \in 1..c_{i}^{f}}(\mu_f - 1)z_{ik}^{fn} =0$ the constraint states that the in-flow
and out-flow are the same, that is, if no VNF is traversed, the
flow remains unchanged. The same result is obtained for all VNF
$f$ such that $\mu_f = 1$ (no compression/decompression).
To the aim of linearizing this constraint, we introduced
variable $\psi_{ik}^{fn}$ (still non-linear representation):
\begin{equation*}
\psi_{ik}^{fn}=(\sum_{j \in N: (j,i) \in A} \phi_{ji}^k) z_{ik}^{fn}
\end{equation*}
The constraints can be linearized using Equations~\eqref{eq:linphi1}-\eqref{eq:linphi3}, with the parameter $M_i$ equal to $\sum_{(j,i) \in A}\gamma_{ji}$, which represents the maximum quantity of flow entering node $i$. If  $(\mu_f - 1)z_{ik}^{fn} =1$ then $\psi_{ik}^{fn}$ representing the flow of demand $k$ entering node $i$
and passing through the copy $n$ of the VNF $f$ (constraint~\eqref{eq:linphi1}-\eqref{eq:linphi2}), otherwise it is zero (constraint~\eqref{eq:linphi3}).
It is now possible to present the new constraints that must
be added to the basic VNF-PR model:
Flow balance for access nodes:
\begin{eqnarray}
&\sum\limits_{j \in N: (i,j) \in A} \phi_{ij}^k - \sum\limits_{j \in N: (j,i) \in A} \phi_{ji}^k = \nonumber\\
& \quad = \left \{
   \begin{array}{c r}
    b^{k}  &\text{if} \quad  i=o_k\\
    0      &\text{otherwise}\\
    -b^{k} \prod\limits_{f \in F : \text{m}_{k}^{f}=1}\mu_f 
    	   &\text{if} \quad i=t_k
   \end{array}
   \right . \forall k \in D, i \in N_a
   \nonumber\\
\label{eq:flowaccess}
\end{eqnarray}
Flow and compression/decompression balance for NFVI nodes and for each demand:
\begin{eqnarray}
\sum\limits_{j \in N: (i,j) \in A} \phi_{ij}^k - \sum\limits_{j \in N: (j,i) \in A} \phi_{ji}^k =& \nonumber\\ 
\sum\limits_{f \in F, n \in 1..c_{i}^{f} } (\mu_f -1) \psi_{ik}^{fn} 
\quad &\forall k \in D, i \in N_v\label{eq:flowdec}
\end{eqnarray}
Coherence between path and flow variables:
\begin{eqnarray}
&\phi_{ij}^k \le b_k^{max} x_{ij}^k \quad \forall k \in D, (i,j) \in A \label{eq:cohexphi1}\\
&\phi_{ij}^k \ge b_k^{min} x_{ij}^k \quad \forall k \in D, (i,j) \in A \label{eq:cohexphi2}
\end{eqnarray}
VNF compression/decompression linearization constraints:
\begin{eqnarray}
&\psi_{ik}^{fn} \le \sum\limits_{j \in N: (j,i) \in A} \phi_{ji}^k +M_i(1-z_{ik}^{fn}) \nonumber\\
& \quad \forall k \in D, i \in N_v, f \in F, n \in 1..c_{i}^{f} \label{eq:linphi1}\\
&\psi_{ik}^{fn} \ge \sum\limits_{j \in N: (j,i) \in A} \phi_{ji}^k +M_i(1-z_{ik}^{fn}) \nonumber\\
& \quad \forall k \in D, i \in N_v, f \in F, n \in 1..c_{i}^{f} \label{eq:linphi2}\\
&\psi_{ik}^{fn} \le M_i z_{ik}^{fn} \nonumber\\
& \quad \forall k \in D, i \in N_v, f \in F, n \in 1..c_{i}^{f} \label{eq:linphi3}
\end{eqnarray}
One compression/decompression VNF per node and demand:
\begin{align}
\sum\limits_{f \in F} \sum\limits_{n \in 1..c_{i}^{f}: \mu_f \neq 1} z_{ik}^{fn} \le 1 \quad \forall k \in D,  \forall i \in N_v  \label{eq:atmostoneVFdemand}
\end{align}
Eq.~\eqref{eq:flowaccess} represents the flow balance for the access nodes.
At destination node the quantity of flows is set equal to the
demand multiplied for all factors of compression of all the
demanded VNFs. Eq.~\eqref{eq:flowdec} represents the flow balance for a
given node that has the possibility of hosting VNFs (NFVI). Eq.~\eqref{eq:cohexphi1}-\eqref{eq:cohexphi2} allow to connect variables $x$ and $\phi$, in such a
way that only and only if arc $(i,j)$ is used by demand $k$,
that is $x_{ij}^k = 1$, then variable $\phi$ can be different from
zero. As the demand passes through VNF that can compress or
decompress the flow, then we can determine upper and lower
bound for the demand that are: $b_k^{max} = b_k \prod_{f \in F: \mu_f \ge 1}$ and $b_k^{min} = b_k \prod_{f \in F: \mu_f \le 1}$\footnote{To avoid these parameters being zero when does not exist any VNF with
$\mu_f \ge 1$ (only decompression), and $\mu_f \le 1$, respectively, the calculation of
the parameter can be modified in 
$b_k^{max} = b_k \max\{1,\prod_{f \in F: \mu_f \ge 1}\}$ and $b_k^{min} = b_k \min\{1,\max\{0,\prod_{f \in F: \mu_f \ge 1}\}\}$, respectively.}. Variables $x$ are still necessary to
impose the isolated cycles elimination and the order in the
VNF chain. The utilization rate constraints must be modified
as follows:
\begin{equation}
\sum_{k \in D} \phi_{ij}^k \le U\gamma_{ij} \quad \forall (i,j) \in A\label{eq:linkutil-cd}\\
\end{equation}

To take into account the combined effect of compression/decompression and VNF latency some modification are
needed.

For the \textit{standard} model, constraints~\eqref{eq:nodelatency-st} are modified as follows:
\begin{eqnarray}
&l_{ik}^{f} \ge g_{j}^{f}(\sum\limits_{d \in D} \psi_{id}^{fn}) - L(1-z_{ik}^{fn}) \nonumber\\ 
&\quad \forall k \in D, i \in N_v, f \in F, n \in 1..c_i^f, j \in 1..n_g \label{eq:nodelatency-st-cd}
\end{eqnarray}
For the \textit{fastpath} model, constraints~\eqref{eq:nodelatency-fp2} are modified as follows:
\begin{equation}
\sum_{d \in D} \psi_{id}^{fn} \le B_{max}^f \quad \forall i \in N_v, f \in F, n \in 1..c_i^f \label{eq:nodelatency-fp2-cd}
\end{equation}
{
\footnotesize
\begin{table} 
\footnotesize
\caption{Applicable constraints to VNF-PR problem variations} 
\begin{tabular}{c|c|c|c}
\hline
&\multicolumn{3}{c}{\textit{Constraints}}\\
\textit{Features}     & routing/location & latency profile & bit-rate\\
\hline
basic        & \eqref{eq:singlepath}, \eqref{eq:linkutil}, \eqref{eq:servercap}-\eqref{eq:order}  & & \\
basic-lat    & \eqref{eq:singlepath}, \eqref{eq:linkutil}, \eqref{eq:servercap}-\eqref{eq:order}  & \eqref{eq:demandlatency}, [\eqref{eq:nodelatency-st} vs \eqref{eq:nodelatency-fp1}-\eqref{eq:nodelatency-fp2}] & \\
basic-cd     & \eqref{eq:singlepath}, \eqref{eq:linkutil-cd}, \eqref{eq:servercap}-\eqref{eq:order}  & & \eqref{eq:flowaccess}-\eqref{eq:atmostoneVFdemand}\\
basic-lat-cd & \eqref{eq:singlepath}, \eqref{eq:linkutil-cd}, \eqref{eq:servercap}-\eqref{eq:order}  & \eqref{eq:demandlatency}, [\eqref{eq:nodelatency-st} vs \eqref{eq:nodelatency-fp1}-\eqref{eq:nodelatency-fp2}] & \eqref{eq:flowaccess}-\eqref{eq:atmostoneVFdemand}\\
\hline
\end{tabular}\label{summary}
\end{table}
}
In Table~\ref{summary} we summarize the different models. In the
first column a short name is used to refer to each model,
in the second column constraints necessary to model routing,
location and resource capacity are reported. In the third and
forth columns we report latency and compression/decompression constraints.

\subsection{Multi-objective math-heuristic resolution}
\label{matheu}

We face a multi-objective problem: minimizing the maximum link utilization, which reflects the ISP-oriented vision to improve the user quality of experience (strictly related to link congestion, especially for real-time services) and minimizing the total virtualization infrastructure cost at the NFVI level, which reflects the aims of the NFVI provider. Such a multi-objective approach makes especially sense when the NFVI provider is a different entity than the ISP. These two objectives are in competition; in fact, to obtain a low utilization, a large number of VNFs must be allocated.

We decided to prioritize the objectives: first we minimize the maximal link utilization ($U$), and then the NFV cost (total number of used CPU). We refer to this as the TE-NFV objective. In practice, we perform a first optimization step to find the best solution accordingly to maximal link utilization ($U^\star$), and then, keeping the best value found in the first step as a parameter (i.e. adding the constraint $U \le U^\star$), we minimize the second objective (NFV cost). In fact, for a given optimal value of the first step, different possible configurations are available to the second step, and a large primary cost reduction can be achieved by this second step without losing with respect to the primary objective (maximum link utilization).

In order to understand the impact of imposing a maximal link utilization on the NFV cost, we decided to study the sensitivity of the second step of optimization on the optimal value $U^\star$. Therefore, we re-optimize the second objective relaxing the constraint on the  maximum link utilization by a parameter $\alpha$, i.e. we used constraint $U \le \alpha +U^\star$ instead of $U \le U^\star$. We increase $\alpha$ step by step, until the value of the NFV cost does not reduce anymore. This value corresponds to first minimize the NFV cost and then the maximum link utilization cost (NFV-TE).

From preliminary tests, we observed that optimizing the complete model is very expensive, and that computational time can be significantly reduced performing a sequence of optimization starting from a basic model to the complete one. The result of each step is used as a starting point for the following one, a so-called \textit{warm-start}, that allows to reduce computational time and/or produce better solutions or gaps (when optimization is stopped before reaching the optimal solution). To be more precise, the sequence of models we optimize is first the basic one (only demand routing, VNF location and capacities are considered), then basic-lat (latency is added) and finally basic-lat-cd (latency and compression/decompression are added), see Section~\ref{netm} and Table~\ref{summary} for the complete description of the models and equations involved.

The most challenging model from an optimization point of view is the last one, basic-lat-cd. For this reason, we need to provide a feasible starting solution (warm start) for this step. To this aim, the previous step, optimizing basic-lat model, must be done with some slightly modification. The compression/decompression feature changes the quantity of flow that traverses the graph, therefore to guarantee that the solution of the second step is feasible for the last one, it is necessary to route a worst case quantity of flow, given by the case that all the VNFs with decompression are already applied to the demand flow\footnote{Of course, this worst case can be improved considering the order of VNFs, when it is known in advance.}.

The NFV objective function results to be computationally more challenging than the TE one. Therefore, for obtaining the optimal solution of the NFV goal, a bisection procedure is used on the number of  allocated VNFs/VMs to guarantee solution optimality, even when in a single step the solver is not able to guarantee it: that is, at each bisection step, if a feasible solution is found, the number of VNF/VM is divided by two, and if no feasible solution exists (proving that the problem is unfeasible results in an easier computational task than finding an optimal solution) then it is doubled.

\subsection{Further model refinements}

The model we provided above can be possibly refined
and customized to meet specific requirements. We list in
the following the possible variants as well as the corresponding
modeling variations.
\begin{itemize}
\item \textit{VNF affinity and anti-affinity rules:} 
due to the privacy, reliability or other reasons, a provider may want impose rules on the placement of certain types of VNF: be placed or not placed on certain servers, be grouped or not grouped together, etc. Such specific VNF placement rules are called affinity and/or anti-affinity rules~\cite{Bouten2016}. 
To extend our model to take them into account, the simplest way is to introduce a new variable representing the presence of a certain type of VNF $f$ on a given node $i$ (we remind the reader that our model allows to have multiple copies of the same type of VNF on the same node). Let call this variable $v_{i}^{f}$, it will be equal to one if a VNF of type $f$ is located on node $i$. To make these variables consistent with already defined variables $y_i^{fn}$, we need to add:
\begin{equation*}
     \sum_{n \in 1..c_{i}^{f}} y_{i}^{nf} \le c_{i}^{f} v_{i}^{f} \quad \forall i \in N_v,f \in F
\end{equation*}

More precisely, common affinity/anti-affinity rules are:

\begin{itemize}
    \item VNF-VNF affinity rules: if two VNFs communicate frequently and should share a host node, we may want to keep the VNFs together in order to reduce traffic across the networks and improve the traffic efficiency. Let $\textrm{AffVV}_{f_1 f_2}$ be a parameter equal to one if $f_1$ and $f_2$ should share the same node. Then: 
    \begin{equation*}
        v_{i}^{f_1} = v_{i}^{f_2} \quad \forall (f_1,f_2) : \textrm{AffVV}_{f_1 f_2} = 1
    \end{equation*}

    \item VNF-Server affinity rules: certain intrusion prevention VNFs should reside in the network edges to guard against worms, viruses, denial-of-service (DoS) traffic and directed attacks. Let $\textrm{AffVS}_{i}^{f}$ be a parameter equal to one if $f$ should be installed on $i$. Then: 
    \begin{equation*}
        v_{i}^{f} = 1 \quad \forall (i,f) : \textrm{AffVS}_{i}^{f} = 1
    \end{equation*}
    
    or restricted to a subset of nodes $S \in N_v$:
    \begin{equation*}
        \sum_{i \in S} v_{i}^{f} = 1 \quad \forall (i,f) : \textrm{AffVS}_{i}^{f} = 1
    \end{equation*}

    \item VNF-VNF anti-affinity rules: it may be required to install multiple instances of a same VNF onto multiple servers in order to improve VNF reliability against failures. Let  $\textrm{AAff}_{f}$ be the anti-affinity parameter; we then impose that at least $nbMin$ nodes host the VNF: 
    \begin{equation*}
      \sum_{n \in N_s} v_{i}^{f} \ge nbMin \quad \forall f : \textrm{AAff}_{f} = 1  
    \end{equation*}
    
    if different VNFs cannot be co-located, let $\textrm{AAffVV}_{f_1 f_2}$ be the anti-affinity parameter and impose: 
    \begin{equation*}
    v_{i}^{f_1} + v_{i}^{f_2} \le 1 \quad \forall (f_1,f_2) : \textrm{AAffVV}_{f_1 f_2} = 1
    \end{equation*}

    \item VNF-Server anti-affinity rules: it may be required to avoid resource-hungry VNFs residing in certain cost-critical servers. Let $\textrm{AAffVS}_{i}^{f}$ be the anti-affinity parameter and impose:
    \begin{equation*}
    v_{i}^{f} = 0 \quad \forall (i,f) : \textrm{AAffVS}_{i}^{f} = 1
     \end{equation*}
\end{itemize}
We can observe that all the constraints that set some variables to one or zero, just reduce the number of variables; 
therefore we can expect that such constraints do not increase the computing time. 
A slightly different condition can be imposed for sharing a VNF among different demands; we refer to that as VNF isolation.


\item \textit{VNF isolation}: 
if the same VNF cannot be shared between
two specific demands, we can add constraints to impose
this condition. It is sufficient to introduce an incompatibility parameter $inc_{k_1 k_2}$, equal to one if demand $k_1$ must
be isolated from demand $k_2$; then we need to add:
\begin{eqnarray*}
z_{ik_1}^{fn} + z_{ik_2}^{fn}\le 1 & \quad \forall i \in N_v,f \in F, \\ 
									&n \in 1..c_i^f, k_1,k_2 \in D
\end{eqnarray*}


\item \textit{Multiple comp./dec. VNFs per NFVI node}: 
to make the presentation simpler, we assumed that in each NFVI node there is at most one VNF that can compress/decompress a flow, i.e. with a factor of compression $\mu_{f} \neq 1$. This assumption can be relaxed using an extended graph in which each node that can host a VNF ($N_v$) is expanded in multiple copies, one for each type of VNF that can be allocated in the node. Otherwise, we can represent all possible combinations of different VNFs allocated to the same node, and adding additional binary variables to represent which combination is chosen.

\item \textit{VNF partial chain ordering}: 
we can observe that partial
order can be imposed with a the same form of constraints
used for total ordering~\eqref{eq:order}, just limiting their number
to existing precedence conditions. It is sufficient to introduce a constraint for each couple of VNFs that has
a precedence relation. More formally, for each demand
$k$, we can introduce a directed acyclic graph $O_k(V_k,P_k)$,
where nodes $V$ represent the set of VNFs that must serve
the demand ($V = \{i \in F : m_k^f = 1\}$), and arcs $P$
represent the order relation between such VNFs, that is
an arc $(i, j) \in P$ if VNF $j$ must be used after VNF $i$.
Then, constraints~\eqref{eq:order} can be rewritten as:
\begin{eqnarray*}
\pi_{jk} \ge \pi_{ik} - (|N_v|+1)(2-w_{ik}^{f_1}-w_{ik}^{f_2}) \\ 
\forall k \in D,\forall i,j \in N_v, f_1,f_2 \in V_k: (f_1,f_2) \in P_k&
\end{eqnarray*}

\item \textit{Additional computing constraints}: 
it can be easily included by tuning existing parameters, as far as computing resource requests can be expressed in an additive way (e.g., for storage).

\item \textit{Load balancing}: 
in the current model, each demand can use a single VNF for each type. The model can be extended to allow per-VNF load balancing. If the load balancing is local to a NFVI cluster, the change in the model is small, in fact it is simply necessary to have some continuous variables taking into account the quantity of demand associated to each VNF. If the load balancing can be between different clusters, then it is necessary to extend the model allowing multiple paths for each demand. However such an extension is expected to largely increase the execution time.

\item \textit{Different VM templates}: 
for the sake of simplicity, differently from~\cite{AddisEtAl2015}, we presented the model considering a one-to-one correspondence between VNF and VM templates (single template). Nevertheless, multiple VM templates can be considered in the model at the price of increasing of one dimension/index all variables indexed on the VNF identifiers. 

\item \textit{Core router as a VNF}: 
if the core routing function is also virtualized, i.e., if the NFVI node and the network router can be considered as a single physical node that runs the core routing function, processing the aggregate traffic independently of the demand, as a VNF, then we need to add a term proportional to InFlow plus OutFlow to \eqref{eq:servercap}:
\begin{align*}
&  \sum_{k \in D} \sum_{f \in F} \sum_{n \in 1..c_{i}^{f}} rr_{r} y_{i}^{fn}\\
&+ \sum_{k \in D} \sum_{j: (i,j) \in A} b_k x_{ij}^k\\
&+ \sum_{k \in D} \sum_{j: (j,i) \in A} b_k x_{ji}^k  \le \Gamma_{ir}  \quad \forall i \in N_v, r \in R 
\end{align*}

If bit-rate compression/decompression is considered, constraint~\eqref{eq:servercap} must be modified as follows:

\begin{align*}
&  \sum_{k \in D} \sum_{f \in F} \sum_{n \in 1..c_{i}^{f}} rr_{r} y_{i}^{fn}\\
&+ \sum_{k \in D} \sum_{j: (i,j) \in A} \phi_{ij}^k\\
&+ \sum_{k \in D} \sum_{j: (j,i) \in A} \phi_{ji}^k  \le \Gamma_{ir}  \quad \forall i \in N_v, r \in R 
\end{align*}

\end{itemize}

\begin{figure}[ht]
\centering 
\includegraphics[width=67mm]{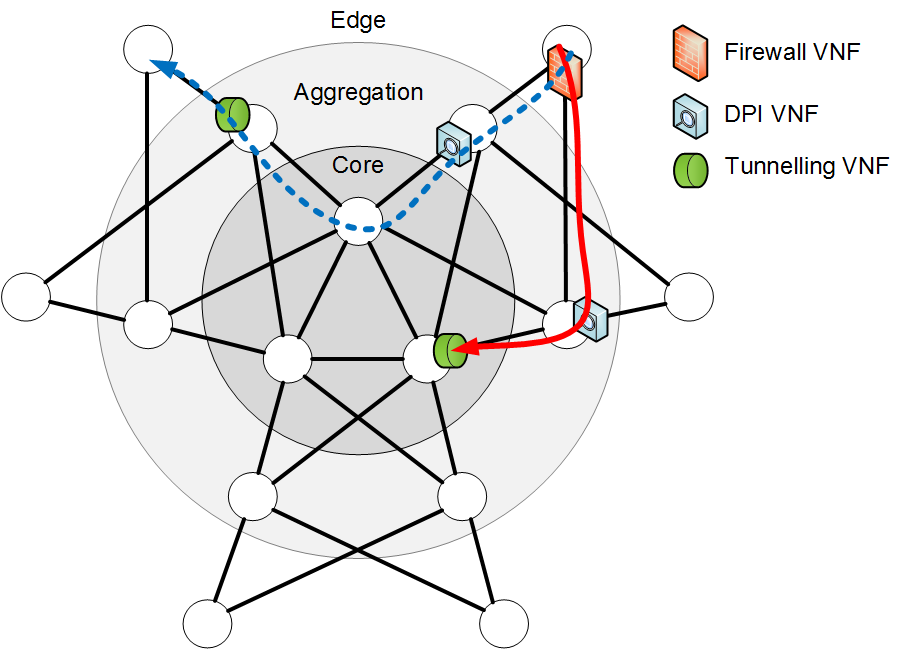}
\caption{Adopted network topology and VNF-PR solution example.}
\label{topo} 
\end{figure}

\section{Results}
\label{results}

Computational results are divided in two main parts: first, in Section~\ref{ssec:results-VNF-PR}, we show results using our VNF-PR algorithm under different case-study choices of demand distribution and different VNF forwarding latency models and values; then, in Section~\ref{ssec:compare}, we present comparison results between our {VNF-PR} algorithm and a VNE algorithm.

We adopted the three-tier topology represented in Fig.~\ref{topo} for computational results. Each edge node is connected to two aggregation nodes, each aggregation node is connected to two core nodes, and core nodes are fully meshed. We consider all the nodes as NFVI nodes that can host VNFs.

We run our tests using two different case-studies for the demand distribution: Internet and Virtual Private Network (VPN). In the Internet case-study (e.g., the flow in red on Fig.~\ref{topo}), the traffic demands are sent by each edge node (e.g., end user) to each core node (e.g., data center), and from each core node to reach each edge node, which means that in this case both edge nodes and core nodes are access nodes (i.e., where demands are generated); while under VPN case-study (e.g., the flow in blue on Fig.~\ref{topo}), the edge nodes send traffic requests to each other, which means that the set of edge nodes corresponds to the set of access nodes. The total number of traffic demands is different for the two case-studies ($36$ for Internet  and $30$ for VPN), but we kept constant the total traffic volume (sum of demands) in the network for the sake of comparison. This is done adjusting the interval of random demand generation $[a,b]$: indeed, the demands are randomly generated with uniform distribution in a given interval $[a,b]$, in such a way that edge demands cannot create any bottleneck at the edge links, i.e., $a = 0.1$ and $b = 0.14$ in the Internet case-study, $a = 0.13$ and $b = 0.17$ in the VPN case-study. These values allow us to keep the total traffic volume at the same level for the two cases. For each case, 10 random demand matrices are considered. 
 
\begin{figure*}[ht]
\centering 
\subfigure[TE objective.]{\includegraphics[height=68mm]{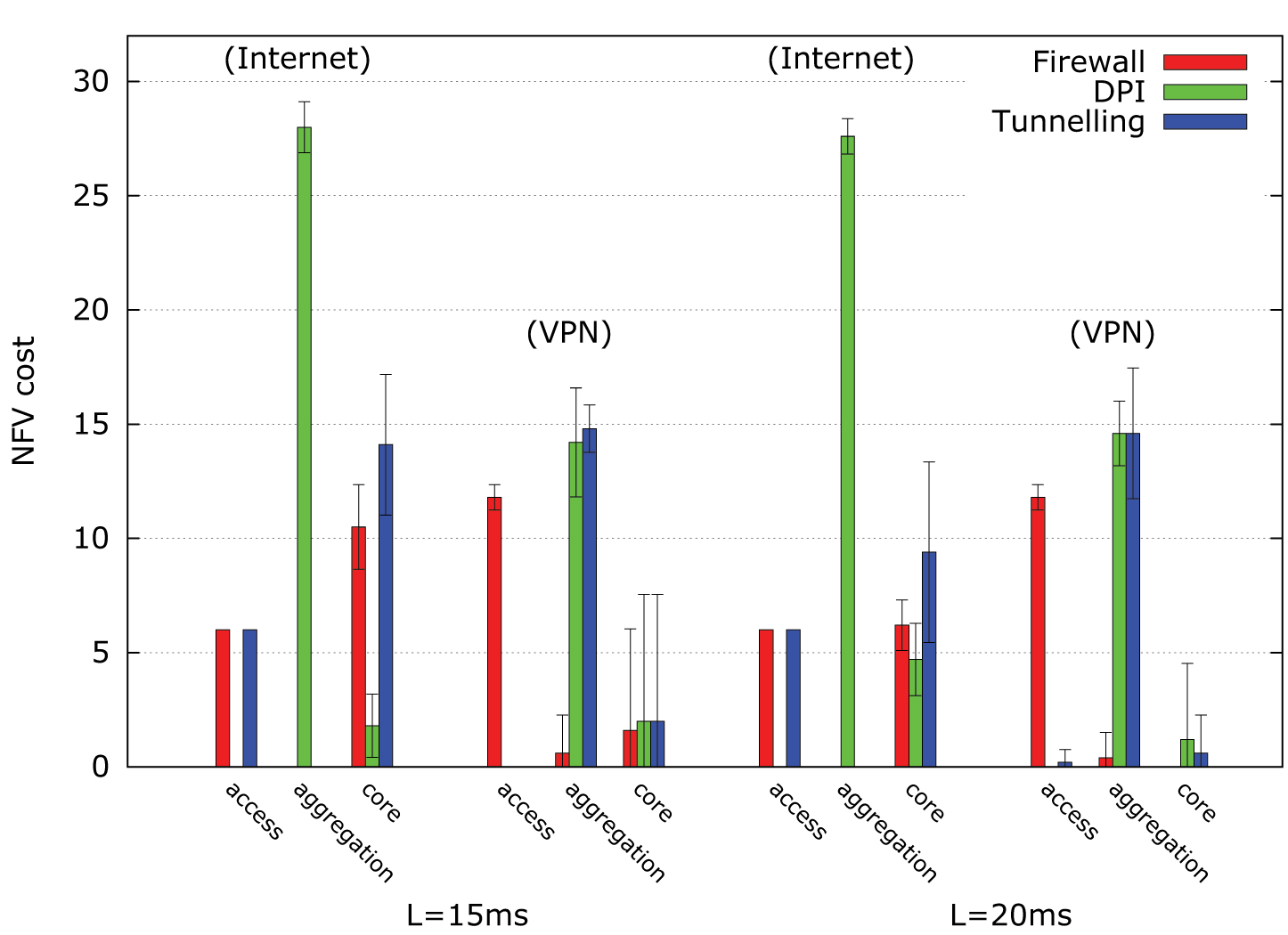}}
\subfigure[TE-NFV objective.]{\includegraphics[height=68mm]{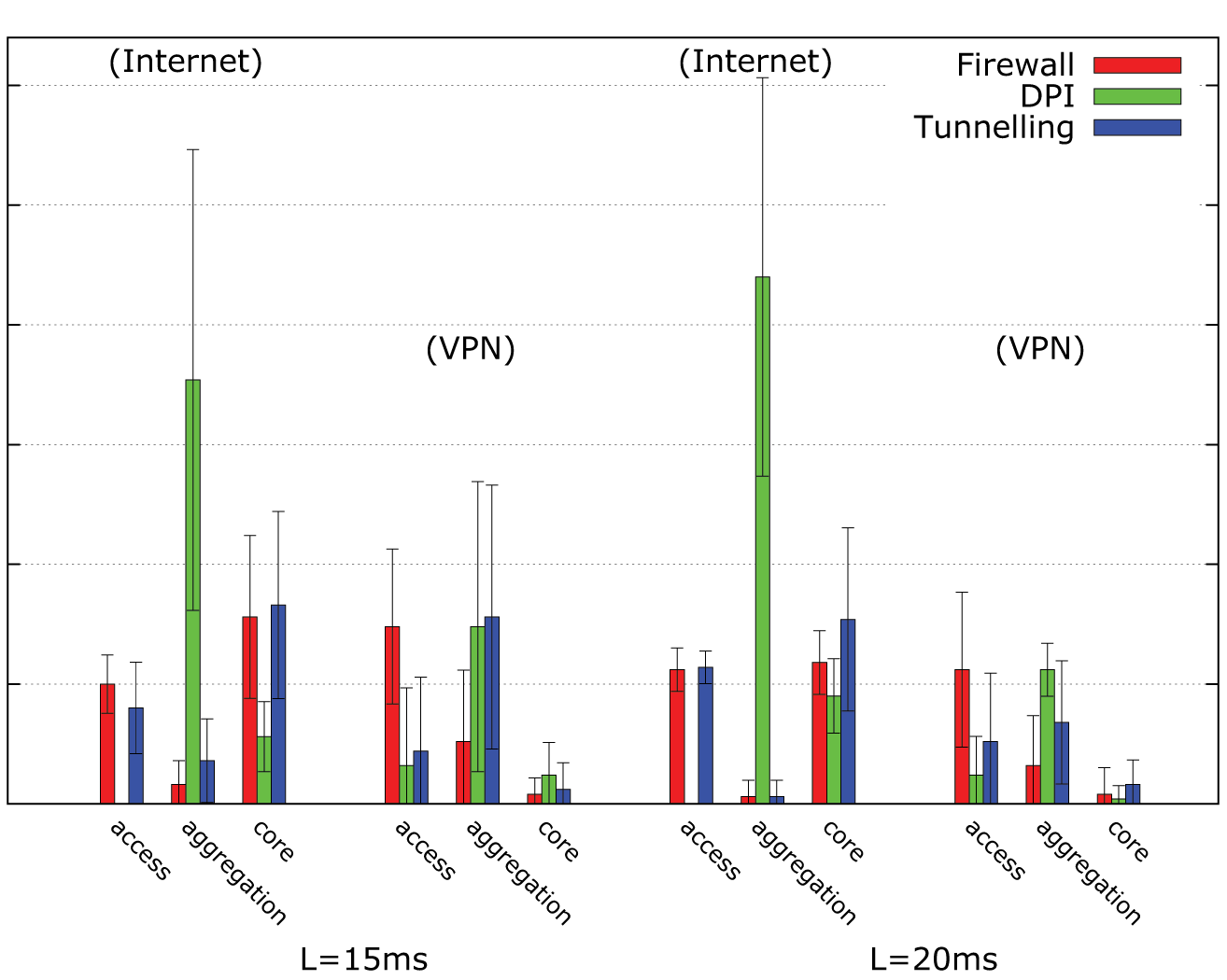}}
\caption{VNF node distribution across NFVI levels (standard case).}
\label{vnfDistrib} 
\end{figure*}

The aggregation links are dimensioned so that there is a risk of link saturation (i.e. link utilization higher than 100\%) if the traffic distribution is not optimized. The core links are such that there is a very low bottleneck risk. Link latencies are set as follows to cope for the different geographical scopes: $1 ms $ for edge links, $ 3 ms $ for aggregation links, and $ 5 ms $ for core links. We use one single VM template requiring 1 CPU and 16 GB of RAM. We run tests setting for the end-to-end latency bound ($ L $) with strict and loose values ($15 ms$ and $20 ms$, resp.). We consider three VNF types per demand: a Firewall VNF (compression, because it blocks traffic), a Deep Packet Inspection (DPI) VNF and a Tunnelling VNF (decompression), with a strict order: Firewall VNF first, then DPI VNF, and finally Tunnelling VNF. NFVI nodes are dimensioned with an increasing capacity from edge to core: 3 CPUs and 40 GB RAM at each edge node, 5 CPUs and 80 GB RAM at each aggregation node and 10 CPUs and 160 GB RAM at core nodes.  
We implemented our VNF-PR algorithm using AMPL and CPLEX 12.6.3.0.

\subsection{General Results}
\label{ssec:results-VNF-PR}

We tested both Internet and VPN case-studies under standard as well as fastpath latency profiles, VNF processing latencies being set as in Fig.~\ref{latency}. 
We limited the execution time to 600s for each basic TE  optimization phase and 800s for the complete TE and NFV optimization phase.

First, we introduce  general considerations from a computational point of view, discussing the quality of the results in terms of optimality and gap of the solutions. Then, we provide a detailed analysis of the structure and properties of the solution in terms of network and system indicators.

In all tests, the worst-case optimality gap with the TE objective is 25\%. Some solutions were proved optimal within 800s (i.e., the limit execution time of the TE optimization phase), others were with an average optimality gap of 15\%. However, these sub-optimal solutions were quickly found (within a few seconds) and could be proved optimal if given more time (around 1 or 2 hours) for almost all the tests. For the NFV objective, it was hard to reach the optimum within the time limit of 800s. The results  depend on the case-study: with VPN demands, under both standard and fastpath latency profiles, we obtained a lower optimality gap and a smaller variation of it than with Internet demands. A possible explanation is the increased number of traffic demands with the Internet case-study, which seems to significantly impact  on the computational effort.  

In the following, we start the analysis of the solutions behavior. We compare the two different demand case-studies (i.e. Internet and VPN) with two points of view: i) what happens when we consider the NFV cost in the objective function instead of TE, and ii) what happens when we make stronger the bound on the end-to-end latency. Then we also compare the behavior with respect to the latency profiles.


\begin{figure}[ht]
\centering 
\includegraphics[width=90mm]{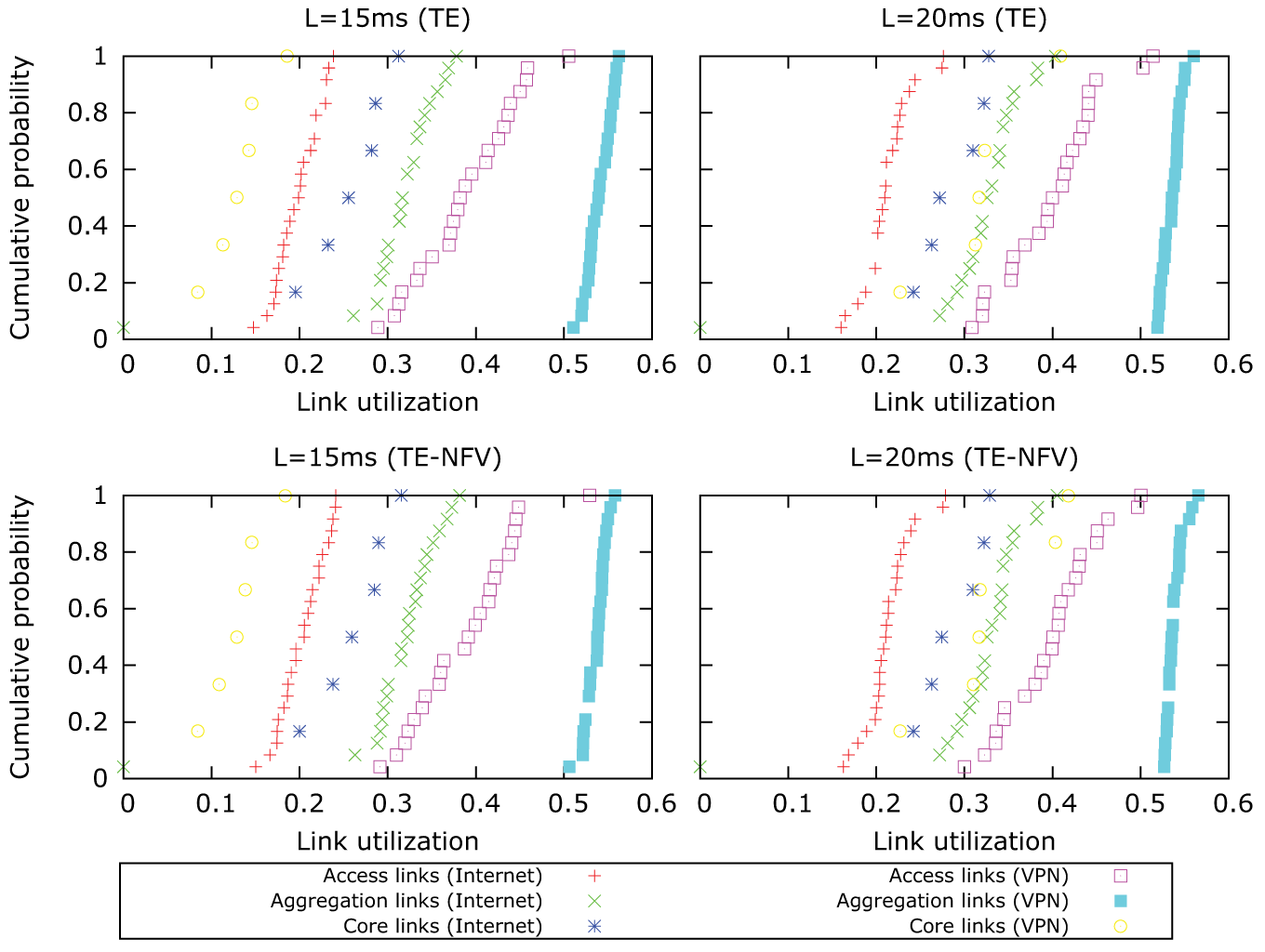}
\caption{Link utilization empirical CDFs (standard case).}
\label{linkUStd4in1} 
\end{figure}

  
\begin{figure}[ht]
\centering 
\includegraphics[width=90mm]{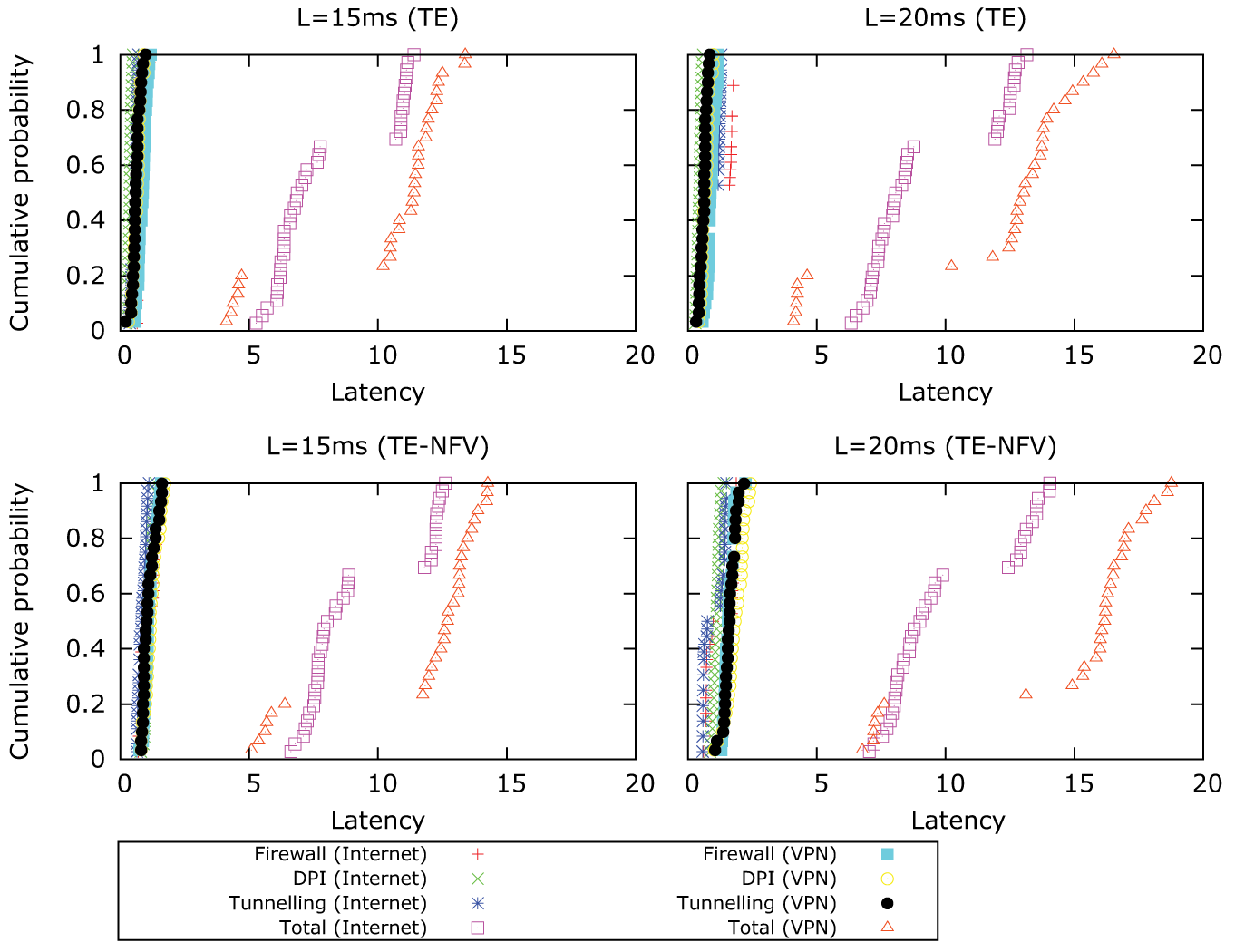}
\caption{Empirical CDFs of latency components (standard case).}
\label{latencyComponentStd4in1} 
\end{figure}

\subsubsection{TE vs. TE-NFV objective}
We analyze the difference between the results with the TE objective and the results with the composite TE-NFV objective.
\begin{itemize}
\item NFVI cost (Fig.~\ref{vnfDistrib}): it is significantly reduced under the TE-NFV objective, and the cost reduction with VPN demands is more significant than with Internet demands, especially with a loose bound on the end-to-end latency ($L=20ms$).
\item Link utilization (Fig.~\ref{linkUStd4in1}): it is not significantly affected by including the NFVI cost minimization in the optimization goal.
\item VNF forwarding latency  (Fig.~\ref{latencyComponentStd4in1}): with both Internet and VPN demands, it increases passing from the TE goal to the TE-NFV one. This suggests that adopting the TE-NFV goal allows a higher level of VNF sharing for both latency bound situations.
\end{itemize}

\subsubsection{Relaxing the TE constraint - sensitivity to maximum link utilization}\label{subsec:sensitivity}

We perform a sensitivity analysis to put in evidence the effect of relaxing the TE objective with respect to the NFV optimal cost, with the goal to further put in evidence the tradeoff between the two objectives. 
With the TE-NFV objective, even if the VNF allocation cost is minimized, a minimum maximum link utilization is guaranteed. What we want to analyze is the impact of the TE bound on the NFV cost objective optimization. To this aim, starting from the TE optimal value, we perform a series of optimization steps of the NFV cost objective function, allowing this bound to be relaxed, increasingly.

Table~\ref{tab:tabAlpha} shows the NFV cost (average of 10 executions) under different limit of maximal link utilization ($U$). 
We calculate the NFV cost under $U = U^\star + \alpha$, with $\alpha$ varying from 0 to 0.4. For both Internet and VPN cases, when $\alpha = 0$, the TE bound ($U$) used for the TE-NFV phase is the $U^\star$ found in TE phase; when $\alpha = 0.4$, the $U$ used for the TE-NFV phase is around 1 (i.e., link saturation reached). The results show that a loose TE bound (link utilization) allows a better TE-NFV solution. For most cases, there is almost no reduction (or no reduction at all) from $\alpha = 0.2$ to $\alpha = 0.4$, which suggests that there exists a ceiling between the TE objective and TE-NFV objective: we can get better utilization of NFV resources (i.e., TE-NFV objective) by allowing relaxed link utilization limit (i.e., TE objective), however this is not always true when we reach the ceiling (e.g., other limits like VM capacity also impact NFV cost). While for case Fastpath Internet $L=20ms$, there is a reduction of NFVI cost with $\alpha$ increasing from $0.2$ to $0.4$. We can observe that for the same case-study  with $L=15$ the best objective found is 31 (both for $\alpha=0.2$ and $0.4$), therefore we can attribute this change in behavior to the not optimality of the solution in the case $L=20$, rather than to a different behavior of the system (we remind the reader that the problem is  computationally very challenging, and we imposed a short time limit).

\begin{table}
\caption{\label{tab:tabAlpha}NFV cost for different TE goal relaxation levels, with TE-NFV optimization.}
\begin{center}
\begin{tabular}{|l||*{3}{c|}}
	\hline
	Instance & $\alpha = 0$ & $\alpha = 0.2$ & $\alpha = 0.4$\\
	\hline
    	&\multicolumn{3}{c|}{ L=15ms }\\
	\hline
    Standard Internet& 48.2 & 25.8 & 24.9\\
	\hline
    Fastpath Internet & 37.125 & 31 & 31\\
	\hline
    Standard VPN & 31 & 28.8 & 28.6\\
	\hline  
    Fastpath VPN & 39.1 & 37.7 & 37.7\\
	\hline  
    	&\multicolumn{3}{c|}{ L=20ms }\\
    \hline
    Standard Internet & 52 & 23.7 & 23.1\\
	\hline
    Fastpath Internet & 38.7 & 34.8 & 31.2\\
	\hline
    Standard VPN & 20.9 & 20.4 & 20\\
	\hline  
    Fastpath VPN & 34.9 & 34.8 & 33.8\\
	\hline     
\end{tabular}
\end{center}
\end{table}

\subsubsection{Sensitivity to the latency bound}
We analyze the impact of the VNF chain latency bound ($L$) on the results.
\begin{itemize}
\item NFV cost (Fig.~\ref{vnfDistrib}): the global NFV cost remains almost constant passing from the weak to the strong latency bound. The total  cost is reduced with VPN demands under both optimization goals with a loose latency bound, especially with the TE-NFV goal. This happens because, with a loose latency bound, the traffic can pass by the links with high latency (e.g. core links) to share more VNFs. On the contrary, there is a small cost increase with Internet demands under TE-NFV goal. Analyzing in a more detailed way the results, we observe that for the Internet case-study, the solver (CPLEX) has more difficulties to reduce the gap. We can deduce that making the latency bound weaker makes the location problem component (locating VNF and NFV goal) predominant with respect to the routing one, and therefore the problem becomes computationally more challenging. This is also confirmed by a noticeable variability in the results, with a larger confidence interval in the number of VNFs with Internet demands passing from strict latency to loose latency bounds.
Moreover, we can see that with VPN demands, there is a higher dependency  to the latency bound than with Internet demands; this happens because in general it is farther to send traffic demands from edge node to edge node than from edge (core) node to core (edge) node, i.e. the end-to-end forwarding path of VPN demands is in general longer than that of Internet demands, which leads to a higher dependency to the latency bound.
\item Link utilization (Fig.~\ref{linkUStd4in1}): in support of the above-mentioned analysis, we can remark that under the loose latency bound, the core links get more utilized with VPN demands. 

\item VNF forwarding latency  (Fig.~\ref{latencyComponentStd4in1}): the same observation can be obtained by looking at end-to-end latency components, the latency of each VNF and the total latency become obviously longer with VPN demands with loose latency bound.
\end{itemize}

These observations confirm the importance of the bound for VNF chaining and placement decisions.

\begin{figure*}[ht]
\centering 
\subfigure[TE objective.]{\includegraphics[height=68mm]{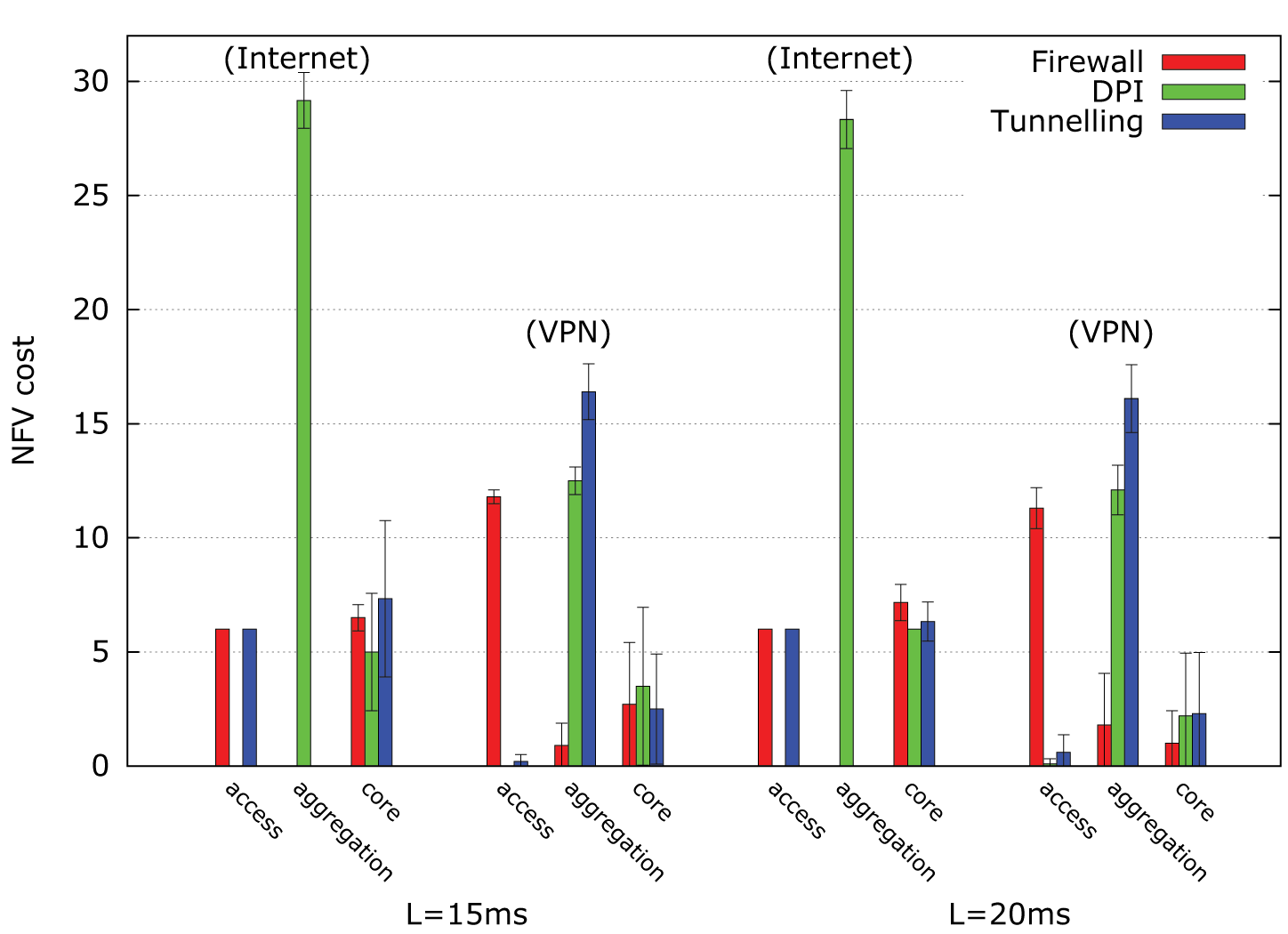}}
\subfigure[TE-NFV objective.]{\includegraphics[height=68mm]{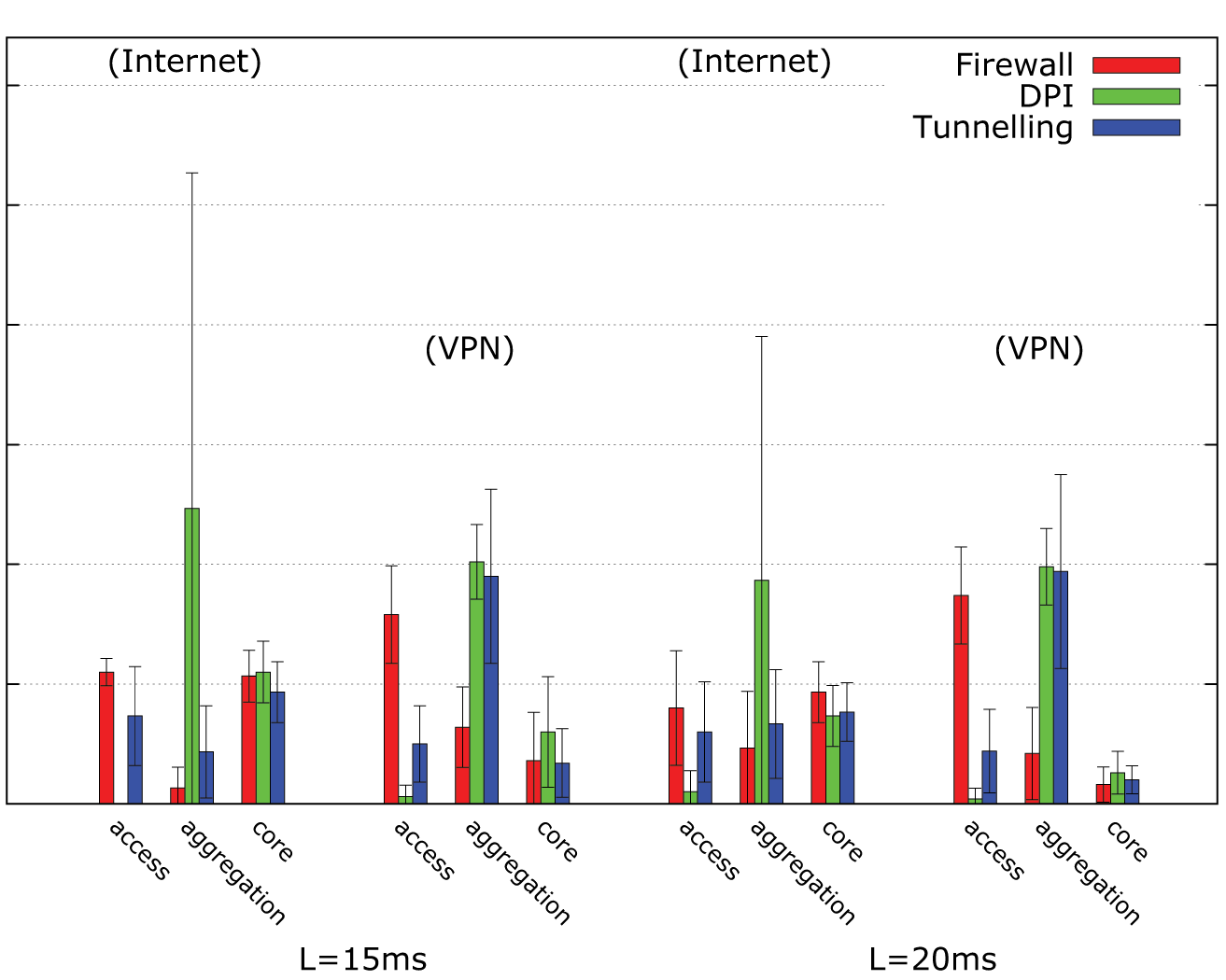}}
\caption{VNF node distribution across NFVI levels (fastpath case).}
\label{vnfDistrib2} 
\end{figure*}

\begin{figure}[t]
\centering 
\includegraphics[width=90mm]{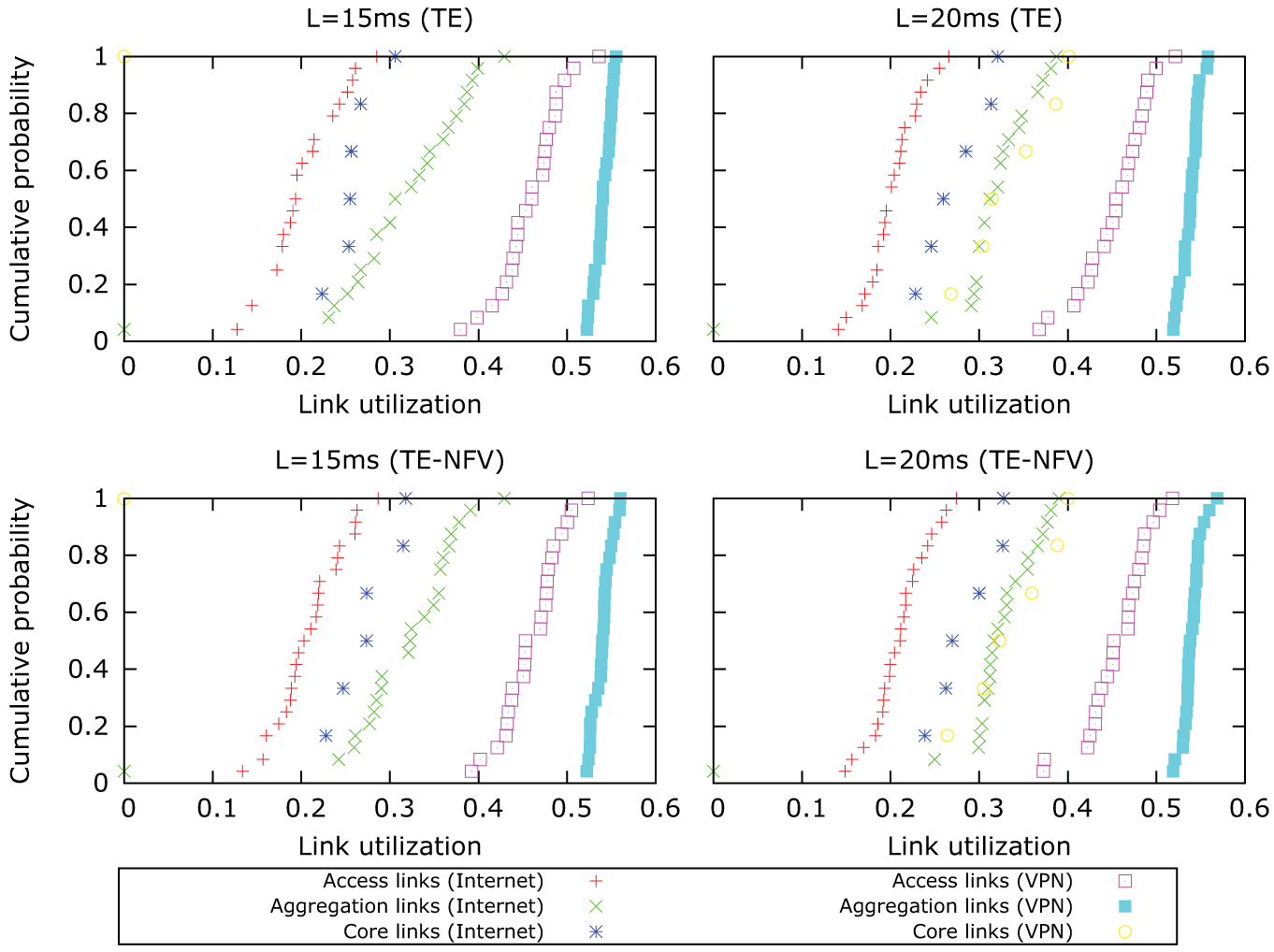}
\caption{Link utilization empirical CDFs (fastpath case).}
\label{linkUFtp4in1} 
\end{figure}

\subsubsection{Standard vs. fastpath VNF switching}
\begin{figure}[ht]
\centering 
\includegraphics[width=90mm]{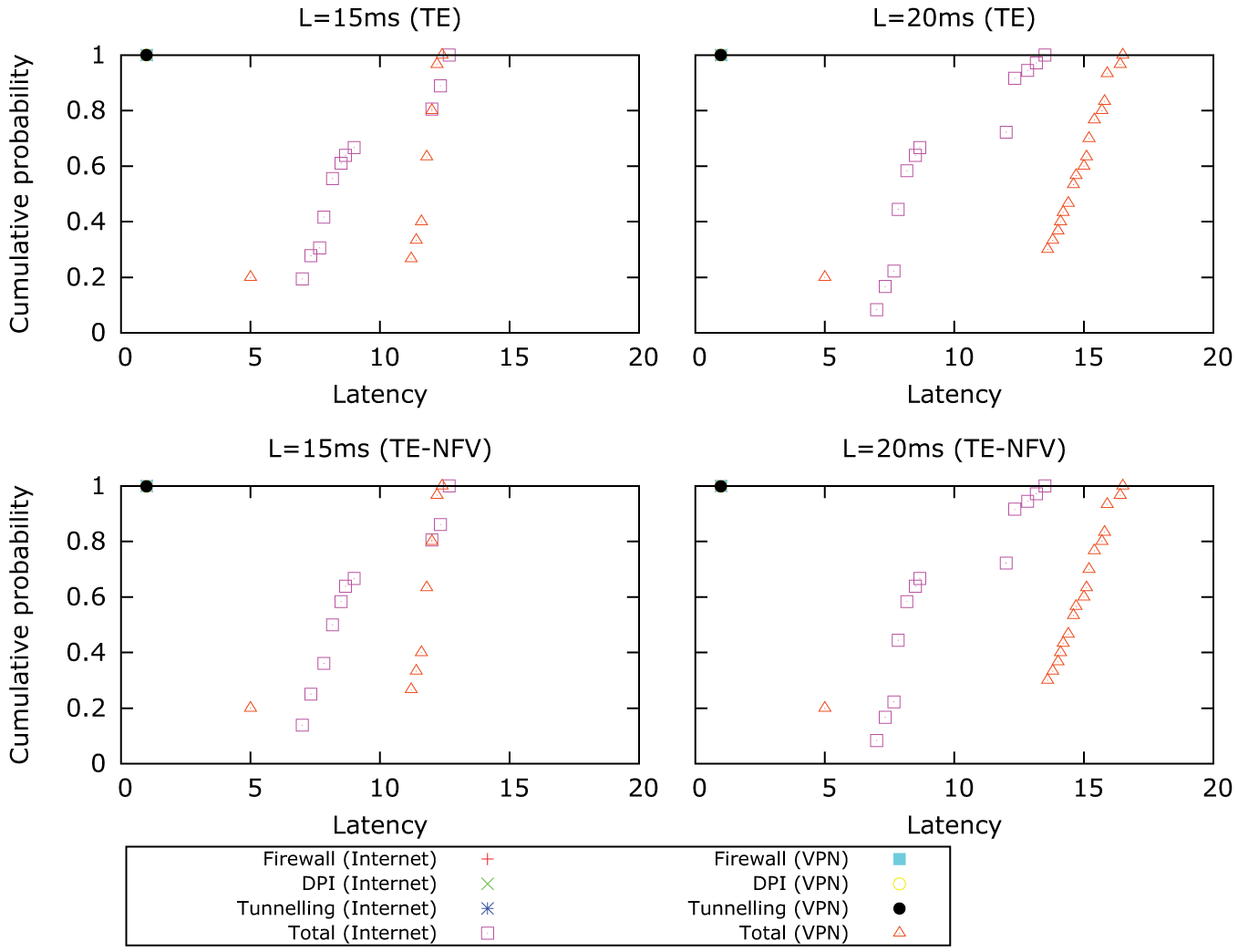}
\caption{Empirical CDFs of latency components (fastpath case).}
\label{latencyComponentFtp4in1} 
\end{figure}

We now compare the results with the standard VNF forwarding latency profile to those with the fastpath profile. 
\begin{itemize}
\item NFV cost (Fig.~\ref{vnfDistrib} vs. Fig.~\ref{vnfDistrib2}): under TE-NFV goal, fastpath VNF forwarding is more expensive than standard forwarding with VPN demands, especially with a loose bound on the end-to-end latency ($L=20ms$), while it is the opposite with Internet demands. This happens because of the maximum traffic bound that is set under the fastpath case and that is not set for the standard case (which however brings to a higher end-to-end latency as confirmed in the last item hereafter).
\item Link utilization (Fig.~\ref{linkUStd4in1} vs. Fig.~\ref{linkUFtp4in1}): no noticeable differences can be mentioned among the two latency profiles.
\item VNF forwarding latency  (Fig.~\ref{latencyComponentStd4in1} vs. Fig.~\ref{latencyComponentFtp4in1}): VNFs are better shared under standard case, this is because of the maximum traffic bound set under the fastpath case.
\end{itemize}

These observations suggest that, performing the optimization in two steps (before TE, and then NFV cost minimization), can allow to reduce significantly the NFVI cost without affecting the link utilization distribution.

\subsection{VNF-PR vs. VNE based approaches}
\label{ssec:compare}

In this section, we compare our VNF-PR approach with the VNE Based (\lq VNE-B\rq) approach, already discussed in the beginning, using the algorithm from~\cite{Raouf2015}, which is open sourced by the authors\footnote{The source code is at \url{https://github.com/srcvirus/middlebox-placement}.}. 

In~\cite{Raouf2015}, a VNE-B modeling approach is proposed for a generic VNF orchestration problem: each traffic demand is considered as a virtual graph (i.e., $G (N,L)$ in~\cite{Raouf2015}, where $N$ is the set of traffic nodes, i.e., switches or VNFs, and $L$ denotes the links between them) to be embedded in the substrate graph represented by switches/routers and NFVI clusters. The mapping of virtualized traffic demands' path onto a physical network is realized by embedding VNFs on physical servers and establishing path for virtual links. The objective considered is the minimization of the overall OPEX (Operational Expenditure) cost: VNF deployment cost, energy cost, traffic forwarding cost and an additional penalty to take into account Service Level Objective (SLO) violations. A weighted sum of the four aforementioned costs is considered as optimization objective. Authors proposed an ILP model, and they presents two different problems, a static one - where demands are know in advance - and a dynamic one - where demands arrive in an online fashion. In our comparison, we focus on the static version of the problem and its proposed solution approach; it is based on a procedure that solves a sequence of ILPs, where, for each iteration, the number of VNFs is limited and the execution time is limited as well. ILP executions are solved with CPLEX (using the callable library). 

Similarly to the NFV cost optimization phase of our VNF-PR approach described  in Sect.~\ref{matheu},  authors in~\cite{Raouf2015} use a dichotomy on the number of VNFs. For the sake of comparison, we adapt our original VNF-PR model to the hypothesis used in~\cite{Raouf2015}: we integrate our adapted VNF-PR ILP model into the same procedure used in~\cite{Raouf2015} (i.e., replacing their ILP model by ours in the procedure) - we refer to this solution as \lq VNF-PR-D\rq. We list the simplifications and adaptations to our model in order to use the same parameters used in the VNE-B approach:
\begin{itemize}
\item We reduced our objective to a single objective: minimizing the overall network operational cost, using the same parameters of the VNE-B approach.   
\item  We considered only the \textit{fastpath} latency regime, i.e., we fixed the VNF forwarding latency.
\item  We discarded compression/decompression aspects, i.e., we adopt our \lq basic-lat\rq \ model.
\item  As the VNE-B approach uses VNF templates, we associated a VNE template to each VNF type according to VNF requested number of vCPUs; for example, a template with capacity of 4 CPU is associated to the VNF requesting 4 CPU.
\item  We added the penalty parameter for each traffic request to take into account SLO violations.
\end{itemize}

\begin{figure}[t]
\centering 
\includegraphics[width=90mm]{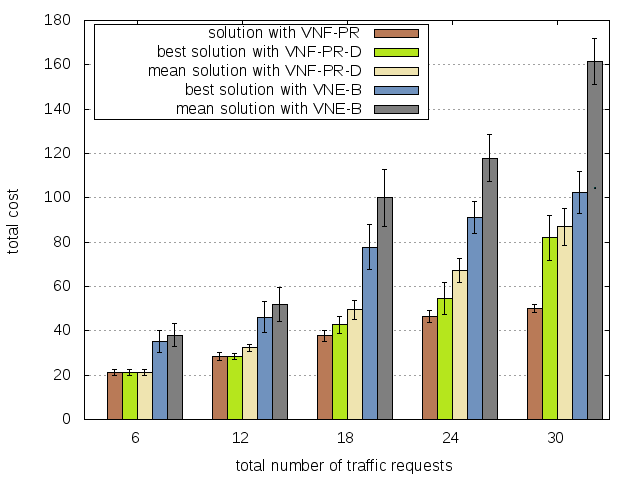}
\caption{Comparison between VNE-PR and VNE based algorithms in terms of objective function.}
\label{vne} 
\end{figure}

As for the test data, they were set according to the simulation setup of~\cite{Raouf2015}, adopting the Internet2 topology (12 switches and 15 links),  setting the same physical link and NFVI server capacities,  using the same VNF specification, VNF requests sequence, etc, and adopting the same cost data. As for the traffic data, we created 5 groups of tests with different sets of traffic requests (6, 12, 18, 24 and 30); for each group, we randomly selected from the traffic matrix set of~\cite{Raouf2015} 10 matrices. Then we tested theses groups of data with the three methods (i.e., VNE-B, VNF-PR and VNF-PR-D). Our algorithms are implemented in AMPL, and the VNE-B algorithm is implemented in C++ using CPLEX for ILP resolution. CPLEX 12.5.1 was used for these tests Fig.~\ref{vne} reports a comparison between our VNF-PR solution and the VNE-B solution in terms of global cost, as a function of the amount of considered demands. As the dichotomy process   could provide a set of feasible solutions found,  we report the best solution as well as the average value of the feasible solution for VNE-B algorithm and VNF-PR-D algorithm. We can notice that the VNF-PR-D is able to find  better solutions compared to the VNE-B algorithm, especially when the amount of traffic demands is small: it can always find optimal solutions. While the VNE-B algorithm is always able to find a feasible solution very fast (i.e., within 10 seconds),  the result can be much worse, and it does not provide optimal solutions even for 6 demands. As for the VNF-PR, it is an exact method to find the optimal solution, therefore, it is in general slow; we run it to find the optimal solution in order to show the gap between the optimal solutions and the feasible solutions. Fig.~\ref{vne} shows that the gap of feasible solutions becomes very important as the traffic requests increase. However, the gap of feasible solutions that was found by VNF-PR-D is smaller than that of VNE-B, which suggests that better problem formulations are needed in order to achieve a trade-off on the execution time and the solution optimality.

\section{Conclusion}
\label{concl}

In this paper we proposed a VNF chaining and placement model, including an algorithm formulated as a mixed integer linear program, which goes beyond recent work in the state of the art. Our model took into consideration specific VNF forwarding modes (standard and fastpath modes), VNF chain ordering constraints, as well as flow bitrate variations; these constraints make the allocation of edge demands over VNF chains unique yet complex. We also mentioned how additional properties being discussed for NFV systems can be integrated in the proposed formulation.
In order to master the time complexity of the resolution algorithm, while considering two different optimization goals -- traffic engineering (TE) goal alone and TE goal combined with NFV infrastructure cost minimization goal (TE-NFV) -- we designed and evaluated a math-heuristic resolution method.
Additionally, we compared our VNF-PR approach to the classical VNE approach often proposed in the literature for NFV orchestration.

We run extensive tests to evaluate our algorithm on a three-tier topology representing an ISP topology. The results showed that the combined TE-NFV objective significantly reduces the number of VNFs in the network compared to the TE objective with almost no impact on the link utilization and on the latency. Moreover we observed that, in addition of different optimization objectives (TE and TE-NFV), the different distributions of traffic demand (Internet and VPN case-studies) and the different VNF types (in terms of function on the bitrate) could lead to different placements of VNF nodes and different VNF chaining paths.

We also quantitatively compared our VNF-PR model to legacy VNE model. The experimental results showed that our VNF-PR algorithm was more stable and close-to-optimum than the VNE solution. The study also showed that the penalty for SLO violation was 0 in almost all the tests performed with VNF-PR approach, while the SLO violation penalty always existed for all the tests with VNE approach; this proved that our VNF-PR algorithm better defines the end-to-end service provisioning (point-to-point source-destination flow routing) problem. Furthermore, the average forwarding cost of all the tests solved with VNE algorithm was about 4 times more expensive than that of VNF-PR algorithm, which indicated that, compared to the VNE solution, there were significantly less redundant traffic forwarding paths when chaining VNFs with VNF-PR approach.

Our future work is to propose new solution algorithms to solve the NFV orchestration problem even more efficiently, taking into account the facility location structure of the problem, which seems to be the most challenging part in the model. Moreover, we envision to adapt our VNF-PR model to be able to meet more specific requirements such as VNF isolation, dynamic VNF orchestration, and the required extension to be run in a batch mode (on a per-demand basis or per groups of demands).

\section*{Acknowledgment}

This work was partially supported by the ANR Reflexion project (\url{http://anr-reflexion.telecom-paristech.fr}, contract nb: ANR-14-CE28-0019), and the French Investissement d’Avenir PSPC (Projet Structurant pour la Competitivité) FED4PMR project.


\bibliography{mybibfile.bib}

\begin{thebibliography}{10}

\bibitem{AddisEtAl2015}
B.~Addis, D.~Belabed, M.~Bouet, and S.~Secci.
\newblock Virtual network functions placement and routing optimization.
\newblock In {\em 2015 IEEE 4th International Conference on Cloud Networking
  (CloudNet)}, pages 171--177, Oct 2015.

\bibitem{chiosi2012network}
M.~Chiosi and et~al.
\newblock Network functions virtualisation: An introduction, benefits,
  enablers, challenges and call for action.
\newblock In {\em {SDN} and {OpenFlow} World Congress}, 2012.

\bibitem{NFVWP}
{ETSI}.
\newblock {Network Functions Virtualization - Introductory White Paper}.
\newblock October 2012.

\bibitem{NFV-commag15}
B.~Han, V.~Gopalakrishnan, L.~Ji, and S.~Lee.
\newblock Network function virtualization: Challenges and opportunities for
  innovations.
\newblock {\em IEEE Communications Magazine}, 53(2):90--97, Feb 2015.

\bibitem{quinn2015network}
M.~Boucadair, C.~Jacquenet, Y.~Jiang, R.~Parker, and K.~Naito.
\newblock Requirements for service function chaining ({SFC}).
\newblock {\em {IETF} Internet-Draft draft-boucadair-sfc-requirements-06},
  2015.

\bibitem{dpdk}
Intel.
\newblock Impact of the {Intel Data Plane Development Kit (Intel DPDK)} on
  packet throughput in virtualized network elements.
\newblock White Paper, 2009.

\bibitem{Fischer2016}
A.~Fischer and H.~de~Meer.
\newblock Generating virtual network embedding problems with guaranteed
  solutions.
\newblock {\em IEEE Transactions on Network and Service Management},
  13(3):504--517, Sept 2016.

\bibitem{NFVGS002}
ETSI.
\newblock {Network Functions Virtualization - Architectural Framework}.
\newblock December 2014.

\bibitem{NFVGS001}
ETSI.
\newblock {Network Functions Virtualization - Use Cases}.
\newblock October 2013.

\bibitem{Basta2014}
Arsany Basta, Wolfgang Kellerer, Marco Hoffmann, Hans~Jochen Morper, and Klaus
  Hoffmann.
\newblock Applying nfv and sdn to lte mobile core gateways, the functions
  placement problem.
\newblock In {\em Proceedings of the 4th Workshop on All Things Cellular:
  Operations, Applications, \&\#38; Challenges}, AllThingsCellular '14, pages
  33--38, New York, NY, USA, 2014. ACM.

\bibitem{Hawilo2014}
H.~Hawilo, A.~Shami, M.~Mirahmadi, and R.~Asal.
\newblock Nfv: state of the art, challenges, and implementation in next
  generation mobile networks (vepc).
\newblock {\em IEEE Network}, 28(6):18--26, Nov 2014.

\bibitem{Wu2015}
J.~Wu, Z.~Zhang, Y.~Hong, and Y.~Wen.
\newblock Cloud radio access network (c-ran): a primer.
\newblock {\em IEEE Network}, 29(1):35--41, Jan 2015.

\bibitem{bobroff2007dynamic}
N.~Bobroff, A.~Kochut, and K.~Beaty.
\newblock Dynamic placement of virtual machines for managing sla violations.
\newblock In {\em 2007 10th IFIP/IEEE International Symposium on Integrated
  Network Management}, pages 119--128, May 2007.

\bibitem{chaisiri2009optimal}
S.~Chaisiri, Bu-Sung Lee, and D.~Niyato.
\newblock Optimal virtual machine placement across multiple cloud providers.
\newblock In {\em 2009 IEEE Asia-Pacific Services Computing Conference
  (APSCC)}, pages 103--110, Dec 2009.

\bibitem{Fischer2013}
A.~Fischer, J.~F. Botero, M.~T. Beck, H.~de~Meer, and X.~Hesselbach.
\newblock Virtual network embedding: A survey.
\newblock {\em IEEE Communications Surveys Tutorials}, 15(4):1888--1906, Fourth
  2013.

\bibitem{Sherry2012}
Justine Sherry and Sylvia Ratnasamy.
\newblock A survey of enterprise middlebox deployments.
\newblock Technical report, {EECS} Department, University of California,
  Berkeley, Feb 2012.

\bibitem{guerzoni2014novel}
R.~Guerzoni, R.~Trivisonno, I.~Vaishnavi, Z.~Despotovic, A.~Hecker, S.~Beker,
  and D.~Soldani.
\newblock A novel approach to virtual networks embedding for sdn management and
  orchestration.
\newblock In {\em 2014 IEEE Network Operations and Management Symposium
  (NOMS)}, pages 1--7, May 2014.

\bibitem{mehraghdam2014specifying}
S.~Mehraghdam, M.~Keller, and H.~Karl.
\newblock Specifying and placing chains of virtual network functions.
\newblock In {\em 2014 IEEE 3rd International Conference on Cloud Networking
  (CloudNet)}, pages 7--13, Oct 2014.

\bibitem{moens2014vnf}
H.~Moens and F.~D. Turck.
\newblock Vnf-p: A model for efficient placement of virtualized network
  functions.
\newblock In {\em 10th International Conference on Network and Service
  Management (CNSM) and Workshop}, pages 418--423, Nov 2014.

\bibitem{Botero2012}
Juan~Felipe Botero, Xavier Hesselbach, Andreas Fischer, and Hermann de~Meer.
\newblock Optimal mapping of virtual networks with hidden hops.
\newblock {\em Telecommunication Systems}, 51(4):273--282, 2012.

\bibitem{Bo2015}
B.~Han, V.~Gopalakrishnan, L.~Ji, and S.~Lee.
\newblock Network function virtualization: Challenges and opportunities for
  innovations.
\newblock {\em IEEE Communications Magazine}, 53(2):90--97, Feb 2015.

\bibitem{Mijumbi2016}
R.~Mijumbi, J.~Serrat, J.~L. Gorricho, N.~Bouten, F.~De Turck, and R.~Boutaba.
\newblock Network function virtualization: State-of-the-art and research
  challenges.
\newblock {\em IEEE Communications Surveys Tutorials}, 18(1):236--262,
  Firstquarter 2016.

\bibitem{netsoftVNFchain}
R.~Mijumbi, J.~Serrat, J.~L. Gorricho, N.~Bouten, F.~De Turck, and S.~Davy.
\newblock Design and evaluation of algorithms for mapping and scheduling of
  virtual network functions.
\newblock In {\em Proceedings of the 2015 1st IEEE Conference on Network
  Softwarization (NetSoft)}, pages 1--9, April 2015.

\bibitem{Cohen2015}
R.~Cohen, L.~Lewin-Eytan, J.~S. Naor, and D.~Raz.
\newblock Near optimal placement of virtual network functions.
\newblock In {\em 2015 IEEE Conference on Computer Communications (INFOCOM)},
  pages 1346--1354, April 2015.

\bibitem{Luizelli2015}
M.~C. Luizelli, L.~R. Bays, L.~S. Buriol, M.~P. Barcellos, and L.~P. Gaspary.
\newblock Piecing together the nfv provisioning puzzle: Efficient placement and
  chaining of virtual network functions.
\newblock In {\em 2015 IFIP/IEEE International Symposium on Integrated Network
  Management (IM)}, pages 98--106, May 2015.

\bibitem{bouetcost}
M.~Bouet, J.~Leguay, and V.~Conan.
\newblock Cost-based placement of vdpi functions in nfv infrastructures.
\newblock In {\em Proceedings of the 2015 1st IEEE Conference on Network
  Softwarization (NetSoft)}, pages 1--9, April 2015.

\bibitem{Kuo2016}
T.~W. Kuo, B.~H. Liou, K.~C.~J. Lin, and M.~J. Tsai.
\newblock Deploying chains of virtual network functions: On the relation
  between link and server usage.
\newblock In {\em IEEE INFOCOM 2016 - The 35th Annual IEEE International
  Conference on Computer Communications}, pages 1--9, April 2016.

\bibitem{Obadia2016}
M.~Obadia, J.~L. Rougier, L.~Iannone, V.~Conan, and M.~Brouet.
\newblock Revisiting nfv orchestration with routing games.
\newblock In {\em 2016 IEEE Conference on Network Function Virtualization and
  Software Defined Networks (NFV-SDN)}, pages 107--113, Nov 2016.

\bibitem{Elias2015}
J.~Elias, F.~Martignon, S.~Paris, and J.~Wang.
\newblock Efficient orchestration mechanisms for congestion mitigation in nfv:
  Models and algorithms.
\newblock {\em IEEE Transactions on Services Computing}, PP(99):1--1, 2015.

\bibitem{Riggio2015}
R.~Riggio, A.~Bradai, T.~Rasheed, J.~Schulz-Zander, S.~Kuklinski, and T.~Ahmed.
\newblock Virtual network functions orchestration in wireless networks.
\newblock In {\em 2015 11th International Conference on Network and Service
  Management (CNSM)}, pages 108--116, Nov 2015.

\bibitem{xia2014network}
M.~Xia, M.~Shirazipour, Y.~Zhang, H.~Green, and A.~Takacs.
\newblock Network function placement for nfv chaining in packet/optical
  datacenters.
\newblock volume~33, pages 1565--1570, April 2015.

\bibitem{Herrera2016}
J.~Gil Herrera and J.~F. Botero.
\newblock Resource allocation in nfv: A comprehensive survey.
\newblock {\em IEEE Transactions on Network and Service Management},
  13(3):518--532, Sept 2016.

\bibitem{dpdk2}
Intel.
\newblock {Network Function Virtualization Packet Processing Performance of
  Virtualized Platforms with Linux* and Intel Architecture}.
\newblock Technical Report, Oct. 2013.

\bibitem{Papadimitriou98}
C.H. Papadimitriou and K.~Steiglitz.
\newblock {\em Combinatorial optimization: algorithms and complexity}.
\newblock Mineola, NY: Dover, 1998.

\bibitem{Bouten2016}
N.~Bouten, M.~Claeys, R.~Mijumbi, J.~Famaey, S.~Latré, and J.~Serrat.
\newblock Semantic validation of affinity constrained service function chain
  requests.
\newblock In {\em 2016 IEEE NetSoft Conference and Workshops (NetSoft)}, pages
  202--210, June 2016.

\bibitem{Raouf2015}
M.~F. Bari, S.~R. Chowdhury, R.~Ahmed, and R.~Boutaba.
\newblock On orchestrating virtual network functions.
\newblock In {\em 2015 11th International Conference on Network and Service
  Management (CNSM)}, pages 50--56, Nov 2015.

\end{thebibliography}
\bibliographystyle{unsrt}
\end{document}